# Evaluation of A Spatial Microsimulation Framework for Small-Area Estimation of Population Health Outcomes Using the Behavioral Risk Factor Surveillance System

Emma Von Hoene[1]*, Aanya Gupta[2], Hamdi Kavak[3], Amira Roess[4], Taylor Anderson[1]

[1]*Department of Geography and Geoinformation Sciences, George Mason University, Fairfax, VA, USA;* evonhoen@gmu.edu, tander6@gmu.edu

[2] *Thomas Jefferson High School for Science and Technology, Alexandria, VA, USA;* aanyaashi@gmail.com

[3] *Department of Computational and Data Sciences, George Mason University, Fairfax, VA, USA;* hkavak@gmu.edu

[4] *Department of Global and Community Health, George Mason University, Fairfax, VA, USA;* aroess@gmu.edu

*\* Corresponding Author*

## Abstract

Public health decision-making, including identifying health inequalities and planning targeted interventions, relies on population health data across multiple geographic scales. However, such ground truth data is often unavailable. To address this, researchers use small area estimation (SAE) techniques to model health outcomes at county and census tract levels. While multilevel regression modeling is commonly applied, spatial microsimulation has emerged as a state-of-the-art capable of generating estimates at the individual level that can be aggregated to the area level. However, applications of microsimulation in health-related SAE remain narrow and its broader potential for producing reliable, multi-outcome health estimates remains underexplored. To address this, we systematically applied and evaluated spatial microsimulation-based SAE across multiple health risk behaviors and health outcomes, settings, and scales. This study introduces the **Spatial Health and Population Estimator (SHAPE)**, a spatial microsimulation framework that applies hierarchical iterative proportional fitting (IPF) to estimate two health risk behaviors and eleven health outcomes across multiple spatial scales. SHAPE was evaluated using county-level direct estimates from the Behavioral Risk Factor Surveillance System (BRFSS) and both county- and census tract–level data from CDC PLACES for New York (2021) and Florida (2019). Results show that SHAPE's SAEs are moderately consistent with BRFSS ($\bar{r} \approx 0.5$), similar to CDC PLACES ($\bar{r} \approx 0.6$), and strongly aligned with CDC PLACES model-based estimates at both county ($\bar{r} \approx 0.8$) and census tract ($\bar{r} \approx 0.7$) levels. SHAPE is an open, reproducible, and transparent framework programmed in R that meets a need for accessible SAE methods in public health. By constructing synthetic populations that preserve joint distributions among individual characteristics, SHAPE enables detailed analysis of health disparities and supports downstream modeling such as microsimulations and agent-based simulations.

## 1. Introduction and Background

The field of population health addresses a wide spectrum of challenges, spanning infectious and chronic diseases to mental health and health risk behaviors such as smoking and alcohol consumption (Sharma et al., 2025). A common barrier to addressing these issues is the lack of ground truth data capturing health outcomes and behaviors at fine geographic scales. This limits both local and national health decision-makers in planning and management efforts, such as identifying health inequalities or targeting interventions where they are most needed (Rahman, 2017; Wang, 2018). To fill this gap, researchers use small area estimation (SAE), a collection of statistical methods that combine survey and geographic data to generate estimates of population-level health outcomes at various spatial scales (RTI International, 2025).

There are numerous methods for generating SAE of health outcomes, which can generally be grouped into two main approaches: direct and indirect model-based estimates (Rahman, 2017). Direct estimates are calculated using only the survey responses from individuals or households sampled within the specified geographic areas (counties, states) to estimate disease prevalence or other population characteristics. These estimates are typically straightforward to compute, less biased, and reflect the collected data without relying on additional assumptions or models (Corral et al., 2022). However, in areas with small or no samples, or with low response rates, direct estimates can be unreliable or impossible to derive (Ghosh & Rao, 1994; Kong & Zhang, 2020). Alternatively, indirect model-based approaches can produce reliable estimates for under-sampled areas (Corral et al., 2022; Whitworth et al., 2017).

There are many methods for obtaining indirect model-based estimates, several which have been applied for health-related SAE (see Rahman, 2017 for an overview). One of the commonly used approaches in the public health domain is multilevel statistical modeling (Smith et al., 2021). Multilevel modeling is a regression-based method that accounts for the hierarchical structure of data, such as individuals nested within households or geographic areas, by modeling effects at both individual and group levels (Leyland & Groenewegen, 2020). This approach has long been used to estimate health behaviors such as participation in cancer screening (e.g., Berkowitz et al., 2019; Liu et al., 2019) and tobacco use (e.g., Cui et al., 2012; Eberth et al., 2018) as well as health outcomes such as diabetes (e.g., Labgold et al., 2024), obesity (e.g., Zgodic et al., 2021; Zhang, 2013), cardiovascular disease (e.g., Congdon, 2009), and Chronic Obstructive Pulmonary Disease (COPD) (e.g., Zhang et al., 2014). This approach has been widely adopted since it can incorporate multiple individual- and area-level factors, includes random effects to capture unexplained variability between small areas, and generates reliable estimates at fine geographic scales in a cost-effective manner (Rahman, 2017; Zhang, 2013).

Since 2016, the U.S. Centers for Disease Control and Prevention (CDC) has set the gold standard for applying multilevel modeling for SAEs in the US. Their widely used product, CDC PLACES, provides estimates of chronic diseases, risk factors and health behaviors at multiple spatial scales (CDC, 2024b). PLACES uses a multilevel regression and post-stratification approach. In the multilevel regression stage, individual-level variables from the Behavioral Risk Factor Surveillance System (BRFSS) survey (CDC, 2025a), including age, sex, race/ethnicity, and education, are combined with county-level poverty measures from the American Community Survey (ACS) (US Census Bureau, 2025) and state- and county-level random effects. The model produces predicted probabilities for each health variable (e.g., diabetes, obesity, cancer), which are then multiplied by population counts from the U.S. Census to estimate prevalence rates (CDC, 2024c; Zhang et al., 2014). These estimates are widely used by the public health community including researchers and state and local health departments.

More recently, microsimulation approaches such as iterative proportional fitting (IPF) (Deming & Stephan, 1940), combinatorial optimization (CO) (Williamson, 2013), and generalized regression reweighting (GREGWT) (Tanton et al., 2011; Whitworth, 2013) have emerged as a state-of-the-art for SAE and are gaining traction in public health research (Rahman, 2017; Schofield et al., 2017). Unlike statistical model-based methods which estimate prevalence per geographic area, microsimulation approaches generate a realistic population of synthetic individuals, each with a geographic location as well as any relevant attributes. Then, the individuals can be aggregated to produce SAE at various geographic scales (Lovelace & Dumont, 2016). While used widely to generate populations that are used as inputs for agent-based models (ABMs) and microsimulations (Chapuis et al., 2022; Li & O'Donoghue, 2012; Zhu & Ferreira Jr., 2014), these methods have been less often applied to generate SAE datasets with health behaviors and outcomes. For example, Hermes and Poulsen (2012) estimated smoking rates in London neighborhoods, Crespo et al. (2020) generated diabetes prevalence at the census zone level in Santiago, Chile, Wiki et al. (2023) assessed quality of life and well-being in Aotearoa, New Zealand, and Abubakar and Cunningham (2023) produced HIV/AIDS indicators for Nigeria. Spatial microsimulation approaches have also been applied to generate estimates for obesity (Edwards & Clarke, 2009; Koh et al., 2018), mental health and alcohol consumption (Riva & Smith, 2012), chronic back pain (Smalley & Edwards, 2024), and COVID-19 vaccine attitudes and uptake (Von Hoene et al., 2025).

Spatial microsimulation-based SAE offers insights that complement statistical approaches like multi-level regression. By constructing synthetic populations of individuals, it preserves the joint distributions and multivariate relationships among individual attributes while aligning with local area constraints, supporting investigation of socio-economic or geographic health disparities (Campbell & Ballas, 2016). Microsimulation preserves complex, context-specific relationships between determinants and health outcomes and behaviors better than regression-based models that rely on parametric assumptions. The method is flexible across geographic scales, enabling both aggregation and disaggregation of individuals as needed. Additionally, individual-level estimates can be used directly for downstream tasks such as microsimulation or ABMs that would support policy experimentation (Rahman, 2017; Schofield et al., 2017). However, existing studies that use microsimulation approaches for SAE are relatively narrow, often applying microsimulation to a single health outcome or behavior. Therefore, there is a need to assess the generalizability of spatial microsimulation approaches for SAE more broadly across multiple public health outcomes, especially in the US.

To address this gap, this study systematically applied and evaluated spatial microsimulation-based SAE across multiple health risk behaviors and health outcomes, settings, and scales. We developed the **Spatial Health and Population Estimator (SHAPE)**, an open and publicly available framework in R that ingests individual-level health surveys (e.g., BRFSS) and spatially aggregate census data and uses the microsimulation approach for SAE. We applied the SHAPE framework to estimate two health risk behaviors and eleven health outcomes at multiple spatial scales. We evaluated the consistency of SHAPE estimates with county-level direct estimates for New York (2021) and Florida (2019) and with CDC PLACES model-based estimates for the same states. Results show that SHAPE produces SAE that are comparable to those from CDC PLACES, suggesting that SHAPE could serve as a publicly available tool for rapidly generating SAE when such data is unavailable. Furthermore, because SHAPE generates individual-level synthetic populations that preserve joint distributions, it also enables analyses of health disparities with reduced ecological bias and supports downstream applications such as microsimulation and ABM. The remainder of the paper is organized as follows: Section 2 details the data and methods, Section 3 reports the results, Section 4 discusses the findings, and Section 5 concludes the study.

## 2. Methods

We develop, apply, and evaluate the SHAPE framework to estimate two health risk behaviors and eleven health outcomes across multiple settings and spatial scales. The estimated health risk behaviors and outcomes were selected to align with the CDC PLACES categories and measures (CDC, 2024a). The two health risk behaviors simulated are **smoking** and **obesity,** which are well-established predictors of chronic health conditions (Stenholm et al., 2016). The eleven health outcomes estimated include **cancer, asthma, high blood pressure, diabetes, COPD, arthritis, high cholesterol, kidney disease, heart disease, depression,** and **stroke**. County-level BRFSS direct estimates are costly and time-consuming to produce, motivating the use of model-based alternatives such as CDC PLACES and SHAPE. However, the limited availability of direct estimates creates challenges for model evaluation. Consequently, this study assesses the consistency of CDC PLACES and SHAPE in areas where county-level direct estimates are publicly available: Florida (2019) and New York (2021).

**Figure 1** illustrates the two-stage hierarchical workflow used by SHAPE to produce SAEs of health behaviors and outcomes. SHAPE employs IPF to align individual-level survey data with spatially aggregated demographic data and generate synthetic local populations. The process operates in two stages, described in detail in Section 2.2. In **Level 1**, SHAPE fits survey data to aggregated demographic data, reproducing the demographic structure of each local population (e.g., county or census tract). Once this structure is established, the **smoking** and **obesity** responses from the survey data are carried over for each synthetic individual. Aggregating these individuals to the county or tract level produces SAEs of smoking and obesity that are consistent with local demographic distributions. In **Level 2**, SHAPE incorporates the Level 1 smoking and obesity SAEs with the sociodemographic spatial data to refit the model and estimate the prevalence of eleven health outcomes. All SHAPE estimates are evaluated for consistency against BRFSS direct estimates and CDC PLACES model-based estimates. The datasets used in this study are described in **Section 2.1**, while the SHAPE methodology, experiments, and validation are detailed in **Section 2.2.** The datasets and R script implementing the SHAPE framework used to generate the SAEs are available at: https://github.com/evonhoene/SHAPE.

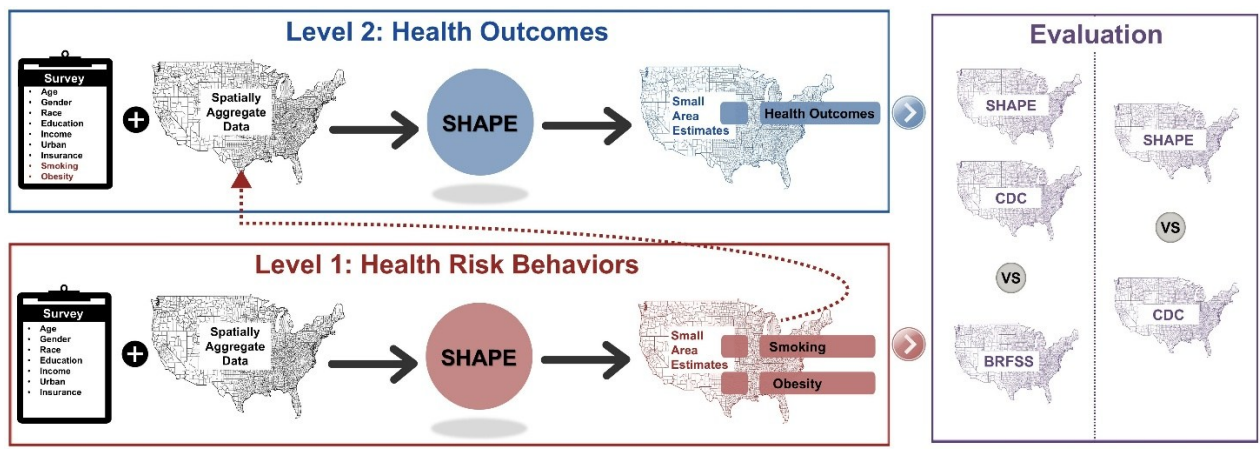

*Figure 1. Schematic overview of the hierarchical IPF framework for generating SAE of health risk behaviors and outcomes using the SHAPE framework, with evaluation conducted using CDC PLACES and BRFSS direct estimates.*

## 2.1. Data

### 2.1.1. Input Data

Given that the IPF-based method used for SHAPE fits disaggregated individual-level survey data to spatially aggregated demographic data with marginal population totals, two main datasets were used to generate the SAEs: *the Behavioral Risk Factor Surveillance System (BRFSS) and the U.S. Census Bureau's American Community Survey (ACS)*. For both datasets, only adults aged 18 and older were included, and we focused on the following sociodemographic predictor variables: age, gender, race, education, income, health insurance status, and urbanicity, all of which have been shown to influence the health risk behaviors and outcomes modeled in this study (Kunnath et al., 2024; National Academies of Sciences et al., 2017).

*Individual-Level Data.* BRFSS is a publicly available national health survey conducted annually by the CDC (CDC, 2025a). It is the largest continuously administered health survey in the world, collecting telephone-based responses from over 400,000 adults (aged 18 years and older) each year across all 50 states, the District of Columbia, and participating U.S. territories. It provides detailed information on respondents' socio-demographics, health behaviors, chronic conditions, and healthcare access. Since this study generated SAEs for Florida (2019) and New York (2021), the 2019 and 2021 BRFSS datasets were used, respectively. The total number of respondents in the dataset was N = 418,268 (2019) and N = 438,693 (2021).

**Stratified and State-Specific Sampling of BRFSS Data.** To manage the large size of the BRFSS datasets and reduce computational cost, the survey was sampled in two ways, as follows:

1. *Stratified Sampled Survey:* Each BRFSS survey was stratified by the seven sociodemographic variables that were used as predictors—age, gender, race, education, income, health insurance status, and urbanicity—to create a sample of approximately 15,000 respondents that accurately reflected the sociodemographic composition of the original survey. In this approach, the population is divided into subgroups (strata) based on shared characteristics, and random samples are drawn proportionally from each stratum. This approach yielded N = 14,968 for the 2019 survey and N = 14,981 for the 2021 survey.
2. *State-Specific Sampled Survey*: Each BRFSS survey was also sampled by state. In other words, survey responses were filtered by the respondent's state of residence, extracting only Florida respondents from the 2019 BRFSS dataset and New York respondents from the 2021 BRFSS dataset. This resulted in N = 10,473 for Florida and N = 20,925 for New York.

For each of these surveys, any responses with missing values for the sociodemographic, health risk behavior, or outcome variables focused on in this study were excluded. Descriptive statistics for the full BRFSS datasets and both sampled surveys for 2019 and 2021 are provided in the ***Appendix, Table 1.*** Additionally for each sampled survey, regression analyses between the sociodemographic predictor variables and each health risk behavior and health outcome modeled in this study are provided in the ***Appendix, Table 2.*** We test the effect of using both sampling strategies—referred to as the Stratified and State-Specific Model experiments—on the resulting SAEs (see Results section).

**Spatially Aggregated Data.** Spatially aggregated data were obtained from the U.S. Census Bureau's ACS (US Census Bureau, 2025), which provides marginal totals for sociodemographic variables. Since the analysis generated estimates at both the county and census tract levels, data were collected for both spatial scales. The 2017–2021 five-year ACS estimates were used for Florida and New York. After removing records with missing or zero values for any of the sociodemographic variables used in this study, the resulting datasets included 67 counties and 5,070 census tracts for Florida, and 62 counties and 5,264 census tracts for New York. Descriptive statistics for each geography and sociodemographic variable are shown in the ***Appendix, Tables 3 and 4.***

### 2.1.2. Data Pre-Processing

To fit the BRFSS individual-level data to the ACS data, both datasets were processed so that the sociodemographic categories were consistent. The variables were categorized as follows: age (18–29, 30–49, 50–64, and 65+), gender (male, female), education (no bachelor's degree, bachelor's degree or higher), race/ethnicity (White, Black, Hispanic, and other), health insurance (with insurance, without insurance), and urbanicity (living in an urban-designated region, non-urban resident). Income categories varied slightly by state to reflect the available BRFSS income brackets, where the 2019 Florida dataset included < $25,000, $25,000–49,999, $50,000–74,999, and ≥ $75,000, while the 2021 New York dataset included < $25,000, $25,000–49,999, $50,000–99,999, and ≥ $100,000. The BRFSS datasets were recoded to match these categories and converted into binary format for consistency. For example, a value of "1" for "age 65+" indicates that the respondent belongs to that category, while "0" indicates otherwise. Each survey record also captured both the selected health risk behaviors and outcomes along with these sociodemographic indicators.

### 2.1.3. Evaluation Data

The evaluation of SHAPE estimates relied on two datasets: county-level BRFSS direct estimates and CDC PLACES model-based estimates at both the county and census tract levels. Both datasets included estimates for the two health risk behaviors and the eleven health outcomes analyzed in this study, and the descriptive statistics for these estimates are presented in the ***Appendix, Tables 5 and 6.***

**BRFSS Direct Estimates.** To evaluate the 2019 Florida estimates, BRFSS direct estimates and their associated confidence intervals (CIs) for all 67 counties were obtained from the Florida Health CHARTS BRFSS dashboard (Florida Department of Health, 2025). In Florida, BRFSS data are collected at the county level every three years in addition to the standard state-level survey. For New York, 2021 county-level direct BRFSS estimates and their associated CIs for 62 counties were obtained from the New York State Department of Health upon request. The department periodically conducts an expanded BRFSS to complement the annual CDC BRFSS survey, aiming to collect county-specific data.

**CDC PLACES Model-Based Estimates.** CDC PLACES estimates were obtained for both counties and census tracts. Florida's estimates came from the 2021 PLACES release, since estimates are derived from 2019 BRFSS data, and included all 67 counties. The same release provided census tract estimates for Florida, including a total of 4,165 tracts. However, because CDC PLACES uses 2015 census tract boundaries while SHAPE estimates rely on 2020 boundaries to align with the 2017 to 2021 five-year estimates we use for spatially aggregate data, only 3,252 tracts could be matched and were used to evaluate SHAPE estimates. Similarly, in New York, the 2023 PLACES release, based on 2021 BRFSS data, included all 62 counties. Of the state's 4,840 census tracts, 4,311 were able to be joined to the SHAPE dataset due to differences in census boundaries.

## 2.2. SHAPE Framework

### 2.2.1. Spatial Microsimulation for SAE

SHAPE relies on the microsimulation technique IPF to generate SAEs, which is a widely adopted approach due to its long-standing presence and reliability in literature, computational efficiency, and its methodological simplicity (Ye et al., 2009). IPF is an algorithm that adjusts the cell values of a contingency table so that they match known marginal totals from external sources, such as spatially aggregated census-based demographic data (e.g., total population aged 65+ across census tracts). Starting with disaggregated individual samples, such as from a survey, IPF iteratively and sequentially reweighs the cell values so that the resulting synthetic population is aligned with both the joint distributions from the survey and the marginal totals from the aggregate data (Pritchard & Miller, 2012).

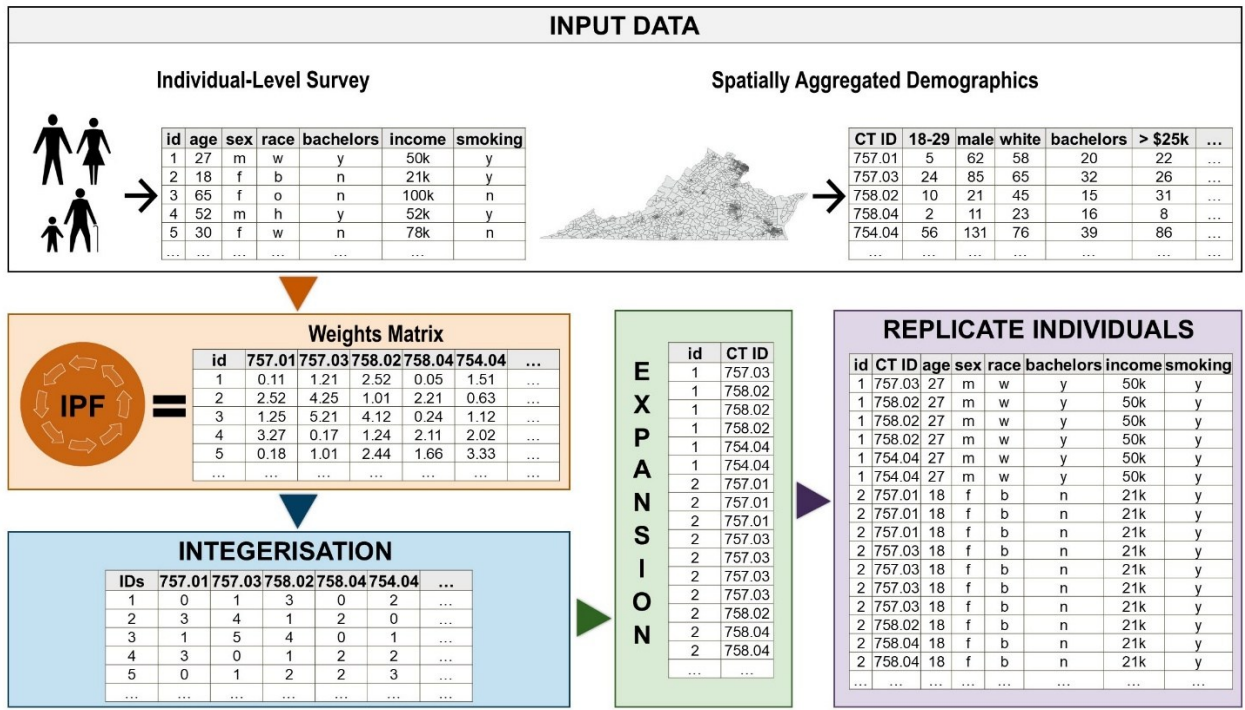

*Figure 2. Overview of the IPF-based population synthesis methodology used in the SHAPE framework, adapted from Von Hoene et al. (2025).*

The IPF approach is summarized in **Figure 2** (adapted from Von Hoene et al., 2025). First, the pre-processed input data, including the sampled individual-level survey and spatially aggregated data, are ingested by the IPF algorithm. IPF assigns a weight to each survey respondent based on how closely their characteristics match the known population distributions for a specific geographic area, such as a county or census tract. Individuals whose characteristics are more common in a given area receive larger weights, while those who are less representative receive smaller weights.

These continuous weights are then converted to whole numbers using the truncate–replicate–sample (TRS) 'integerisation' method (Lovelace & Ballas, 2013). With this procedure, each weight is first truncated to an integer, determining how many times that survey respondent should initially be replicated in the synthetic population. Next, any remaining decimal portions of the weights are used to probabilistically sample additional individuals so that the total population size exactly matches the target population for each geographic zone. This process is repeated for every county or census tract, producing a count of how many copies of each survey respondent are needed to realistically represent the population in that area.

Using these counts, the expansion stage next generates the synthetic dataset, where each record represents an individual from the survey along with their assigned geographic zone. In the final step, any additional attributes from the original input health survey are linked to each replicated individual; Because each record retains the unique survey ID from the input data, any selected variables—such as sociodemographic characteristics, health behaviors, or chronic conditions—can be joined to the synthetic dataset. The result is an individual-level dataset with a total number of agents matching the population from the spatially aggregated data. Ultimately, this dataset can then be aggregated by each agent's home county or census tract to produce health SAEs.

As described above, this process is implemented on two levels (see Figure 1). In Level 1, two health risk behaviors are estimated and used as inputs for the Level 2 estimation of eleven health outcomes. The result is two synthetic populations, consisting of 17,078,449 agents for Florida and 15,897,349 agents for New York, all aged 18 and over. Each agent includes characteristics related to sociodemographic, health behaviors, and health outcomes, along with a home county or census tract. Then, the synthetic populations are aggregated to census tract and county level, and area-level prevalence for health behaviors and outcomes are estimated. This methodology was used to test two experimental setups—the Stratified Model and the State-Specific Model—each based on the different sampled surveys described in Section 2.1.1.

### *2.2.2. Evaluation against BRFSS Direct estimates and CDC PLACES*

To assess the performance of SHAPE in generating SAEs for the two health risk behaviors and eleven health outcomes, we evaluated results from both experiments: the Stratified and State-Specific Models. Since SHAPE employs a hierarchical IPF process, where level one generates estimates for the two health risk behaviors, and level two uses these estimates to generate the eleven health outcomes (see Figure 1), estimates from both levels were evaluated. For each experiment, evaluation consisted of two types of comparisons First, SHAPE and CDC PLACES estimates for Florida (2019) and New York (2021) were compared against BRFSS direct estimates to evaluate consistency using three metrics: Pearson's correlation coefficient (*r*), mean absolute error (MAE), and CI coverage (i.e., the proportion of estimates falling within the BRFSS direct estimates' CIs), as well as spatial heterogeneity. This comparison was conducted only at the county level, given the availability of BRFSS direct estimates. Second, SHAPE estimates were directly compared with CDC PLACES estimates for both states to assess consistency using Pearson's r, MAE, and spatial heterogeneity.

### *2.2.3. Composite Scores and Model Ranking*

To better understand which SHAPE-generated estimates can be used with greater confidence for policy purposes, and which are more suitable for general estimation, we conducted an additional ranking analysis that summarizes model consistency with direct estimates across health behaviors and outcomes for the three models: SHAPE's Stratified Model, SHAPE's State-Specific Model, and CDC PLACES. We first calculated the average value for each of the three evaluation metrics: Pearson's *r*, standardized MAE, and CI coverage across all thirteen health measures, three models, and two states (Florida and New York). The MAE values were standardized using the minimum and maximum values across all models and then inverted (1 – standardized MAE) so that higher values indicated better performance. Next, for each health measure (e.g., smoking, or diabetes), we calculated the deviation of its performance metric from the overall mean. For example, if Florida's smoking *r* value was 0.72 and the average *r* value was 0.58, the deviation was 0.14. Finally, we summed the three deviation scores for *r*, standardized MAE, and CI coverage for each health measure within each model to produce a composite ranking score, with higher positive values indicating that the model tends to be more consistent with direct-estimates than on average and higher negative values indicating that the model is less consistent than on average.

## 3. Results

Recall that SHAPE employs a hierarchical IPF process. In Level 1, survey data are fitted to sociodemographic data from the US Census to generate SAEs for two *health risk behaviors* (i.e., smoking, obesity). In Level 2, survey data are fitted to demographic data and the estimated health risk behaviors to generate SAEs for eleven *health outcomes* (e.g., cancer, diabetes, asthma) (see **Figure 1** ). Here, we present the evaluation results of the SAEs for the *eleven health outcomes* (Level 2, see **Tables 5 and 6 in the Appendix** for the descriptive statistics of SHAPE's estimates). The evaluation results for the Level 1 *health risk behaviors* are included in Section E of the ***Appendix, Tables 7 and 8.***

### 3.1. SHAPE and CDC PLACES vs. Direct Estimates

We first compare county-level SHAPE and CDC PLACES estimates of eleven health outcomes for Florida (2019) and New York (2021) against county-level direct estimates. **Figure 3** summarizes the results for Florida (2019) including:

(Panel A) Pearson's *r* values capturing the correlation between model-based estimates and direct estimates, (Panel B) MAE between model-based estimates and direct estimates, and (Panel C) CI coverage, or the proportion of county estimates falling within direct estimates' 95% CIs. Corresponding results for New York (2021) are shown in **Figure 4**, with panels organized in the same way. Tabular values for each metric are provided in ***Tables 9 and 10 in the Appendix.***

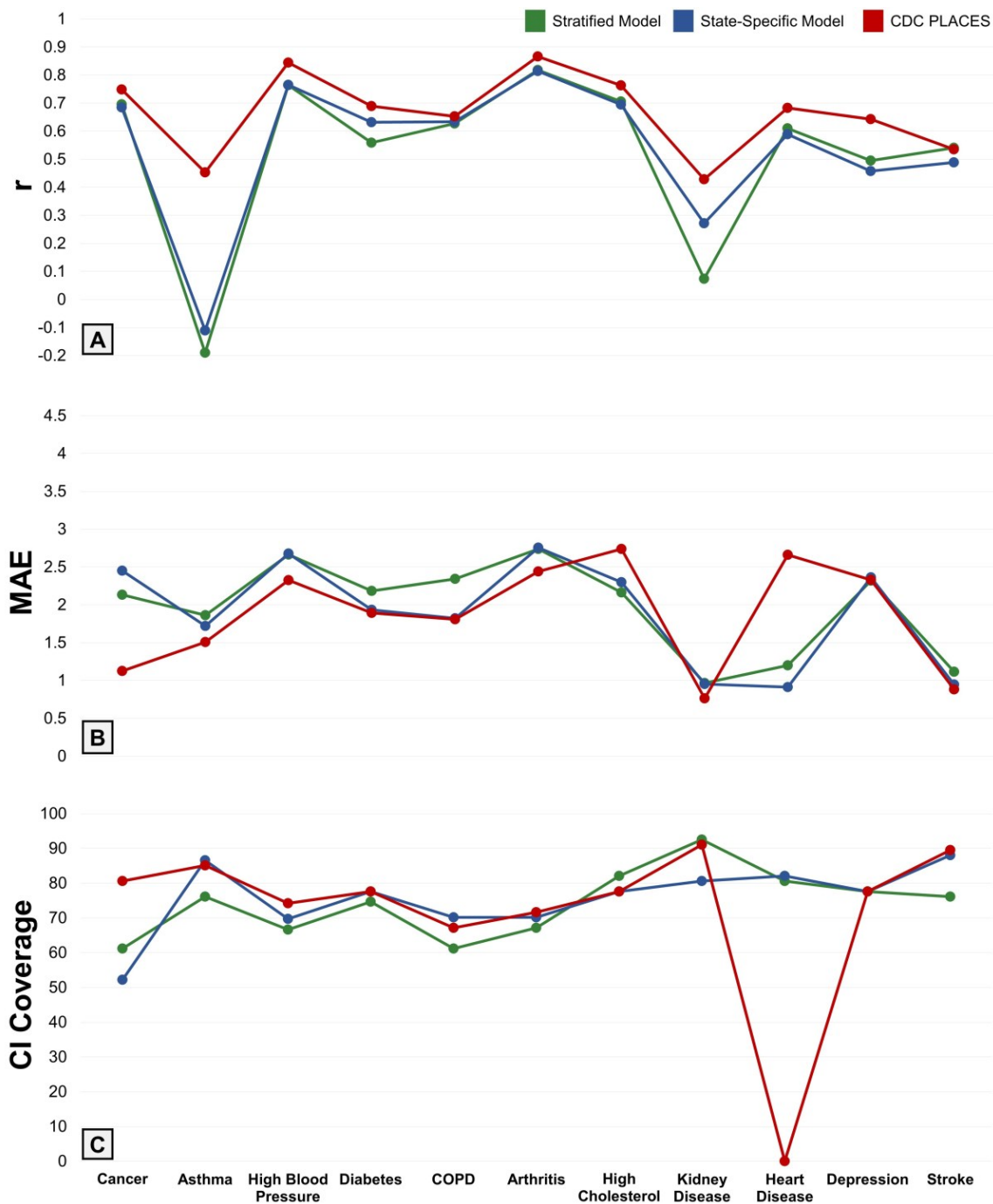

*Figure 3. Line charts evaluating estimates from SHAPE models and CDC PLACES against direct estimates across health outcomes in **Florida, 2019**, using the following metrics: A) R-squared ($R^2$), B) Mean Absolute Error (MAE), and C) Percentage of counties falling within the direct estimates' confidence intervals (CIs).*

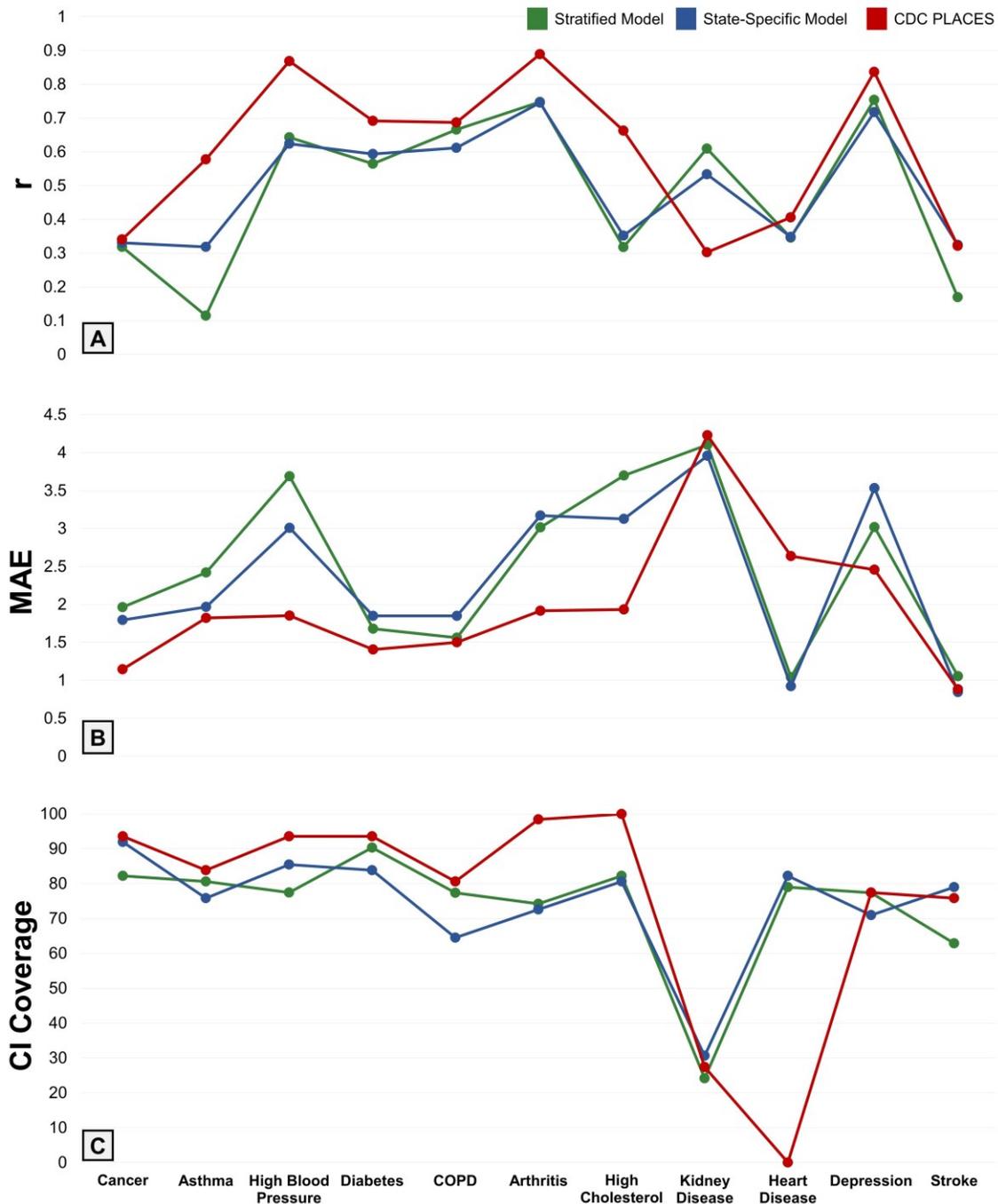

*Figure 4. Line charts evaluating estimates from SHAPE models and CDC PLACES against direct estimates across health outcomes in **New York, 2021**, using the following metrics: A) Pearson's correlation coefficient (r), B) Mean Absolute Error (MAE), and C) Percentage of counties falling within the direct estimates' 95% confidence intervals (CIs).*

### 3.1.1. Consistency with Direct Estimates

In general, when CDC PLACES estimates aligned more closely with direct estimates, SHAPE's estimates did as well. On average, across the eleven health outcomes, SHAPE's Stratified Model showed stronger consistency with direct estimates in Florida ($\bar{r}$ = 0.518, average MAE = 1.972, average CI coverage = 74.175%) than in New York ($\bar{r}$ = 0.478, average MAE = 2.478, average CI coverage = 73.458%). The State-Specific Model performed slightly better than the Stratified Model in both states, Florida ($\bar{r}$ = 0.539, average MAE = 1.894, average CI coverage = 75.672%) and New York ($\bar{r}$ = 0.500, average MAE = 2.366, average CI coverage = 74.342%), with Florida estimates again showing higher overall consistency. As expected, CDC PLACES demonstrated the strongest consistency with direct estimates overall (Florida: $\bar{r}$ = 0.664, average MAE = 1.862, average CI coverage = 72.013%; New York: $\bar{r}$ = 0.625, average MAE = 1.981, average CI coverage = 74.928%).

***Pearson's Correlation.*** In Florida, SHAPE achieved high correlations with direct estimates (r > 0.7) for conditions such as arthritis, high blood pressure, and high cholesterol **(Figure 3A)**. The weakest correlations were observed for asthma, kidney disease, and depression, though CDC PLACES were also poorly correlated to direct estimates for these outcomes. Notably, SHAPE's estimates for asthma in Florida produced the only negative coefficients (Stratified Model = –0.19; State-Specific Model = –0.11), compared to a positive moderate correlation for CDC PLACES ($\bar{r}$ = 0.453). Despite the negative correlation, more than 70% of SHAPE's estimates for asthma fall within the direct estimate's CIs. In New York, SHAPE correlations to direct estimates were generally moderate to strong (r > 0.5) for high blood pressure, diabetes, COPD, arthritis, kidney disease, and depression **(Figure 3A)**. Consistent with CDC PLACES, SHAPE estimates for stroke and heart disease showed lower correlations (r ≈ 0.3), although with relatively small errors (MAE ≈ 1.0).

***Mean Absolute Error.*** Across both states, SHAPE produced estimates with relatively low MAE against direct estimates, indicating close alignment. In Florida, SHAPE achieved MAE values below 2.0 for outcomes including asthma, kidney disease, heart disease, and stroke **(Figure 3B)**. In New York, SHAPE achieved MAE values under 2.0 for multiple outcomes, such as cancer, asthma, COPD, heart disease, and stroke **(Figure 3B)**. Overall, both SHAPE and CDC PLACES had comparable levels of error for most health outcomes, with some exceptions. For Florida, CDC PLACES had a lower MAE for cancer while SHAPE had a lower MAE for high cholesterol and heart disease. For New York, CDC PLACES had a lower MAE for cancer, arthritis, and high cholesterol while SHAPE had a lower MAE for heart disease.

***Confidence Interval Coverage.*** The BRFSS direct estimates for each county include 95% CIs, representing sampling uncertainty. For each outcome, we calculated the percentage of SHAPE county estimates that fall within these intervals. For most health outcomes, approximately 75% of SHAPE county estimates fell within the BRFSS CIs, with the exception of four outcomes in Florida (cancer, high blood pressure, COPD, and arthritis) **(Figure 2C)** and two in New York (arthritis and kidney disease) **(Figure 3C)**. CDC PLACES estimates generally followed a similar trend, but zero of their heart disease estimates fell within the BRFSS CIs for both Florida and New York.

### 3.1.2. Spatial Heterogeneity

As an example, **Figure 5** maps the county-level estimates of COPD prevalence in Florida (2019) and New York (2021) generated from three sources: SHAPE's Stratified Model (panel A, D), BRFSS direct estimates (panel B, E), and CDC PLACES (panel C, F). Across both states, the two model-based estimates capture similar spatial patterns, with lower prevalence in southern Florida and western New York. This visual consistency suggests that both SHAPE and CDC PLACES capture the general geographic distribution of COPD observed in the direct estimates.

Despite broad agreement, the SHAPE estimates **(Figure 4, panel A, D)** appear smoother and show fewer extreme highs and lows, which explains the lower correlation with the direct estimates. This smoothing occurs because IPF does not directly model health outcomes. Instead, it reconstructs each county's demographic composition and assumes that the relationship between demographics and outcomes is consistent across areas. Counties with similar demographic profiles therefore yield similar estimated prevalences. Furthermore, IPF-based estimates are not influenced by random sampling variability from small BRFSS samples, resulting in smoother and more stable spatial patterns. Unlike IPF, CDC PLACES **(Figure 4, panel C, F)** use a hierarchical regression framework that allows for more local variation by considering local contextual factors such as environment, access to care, and community health behaviors, explaining the better correlation with direct estimates. It is important to note, however, that the direct estimates themselves are based on limited survey samples and are more susceptible to random noise that can produce spurious outliers **(Figure 4, panel B, E).**

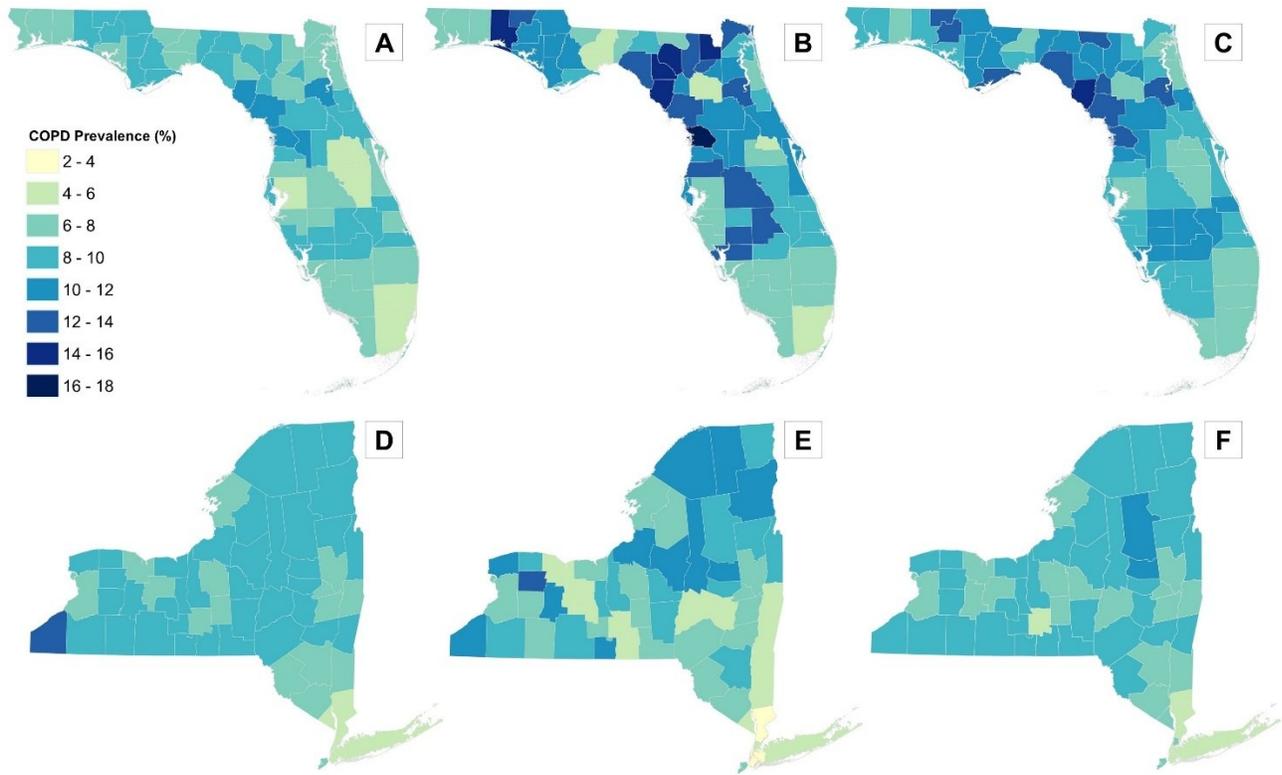

*Figure 5. County-level maps of COPD prevalence estimates from the following sources: A) SHAPE's Stratified Model, Florida 2019; B) BRFSS direct estimates, Florida 2019; C) CDC PLACES estimates, Florida 2019; D) SHAPE's Stratified Model, New York 2021; E) BRFSS direct estimates, New York 2021; F) CDC PLACES estimates, New York 2021.*

### 3.2. SHAPE vs. CDC PLACES

Direct estimates are only available at the county level and are very limited. Given that CDC PLACES is the gold standard for SAE with estimates at multiple scales, we next compare SHAPE estimates with CDC PLACES at both county and census tract level. **Figure 6** summarizes the results for Florida (2019) at both spatial scales, with panel A displaying Pearson's *r* values and panel B showing MAE for each model. Corresponding results for New York (2021) are shown in **Figure 7**, organized in the same format. Tabular values for each metric are provided in ***Tables 11 and 12 in the Appendix.***

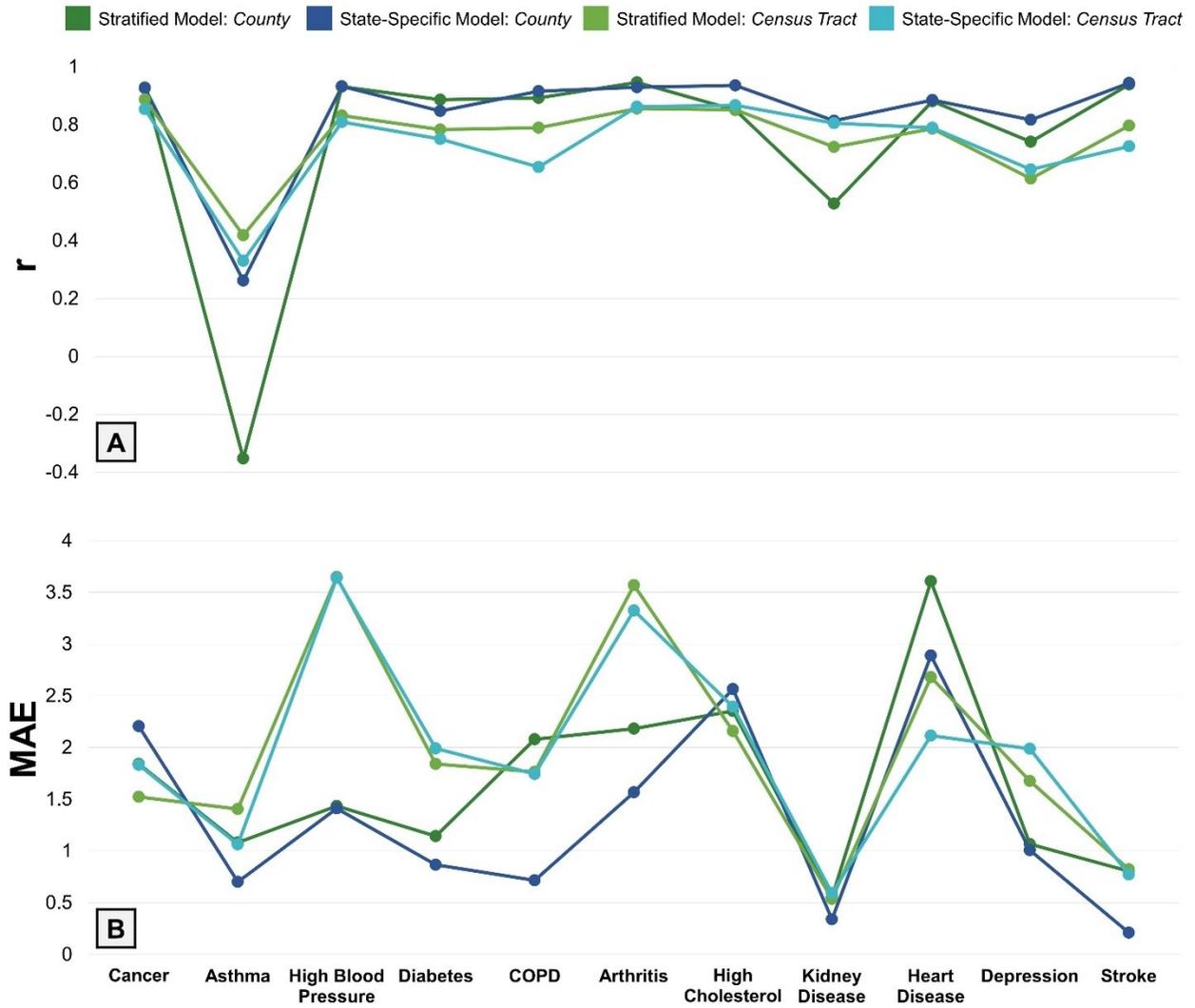

*Figure 6. Line charts evaluating both county and census tract level estimates from SHAPE models against CDC PLACES across health outcomes in **Florida, 2019**, using the following metrics: A) Pearson's correlation coefficient (r), and B) Mean Absolute Error (MAE).*

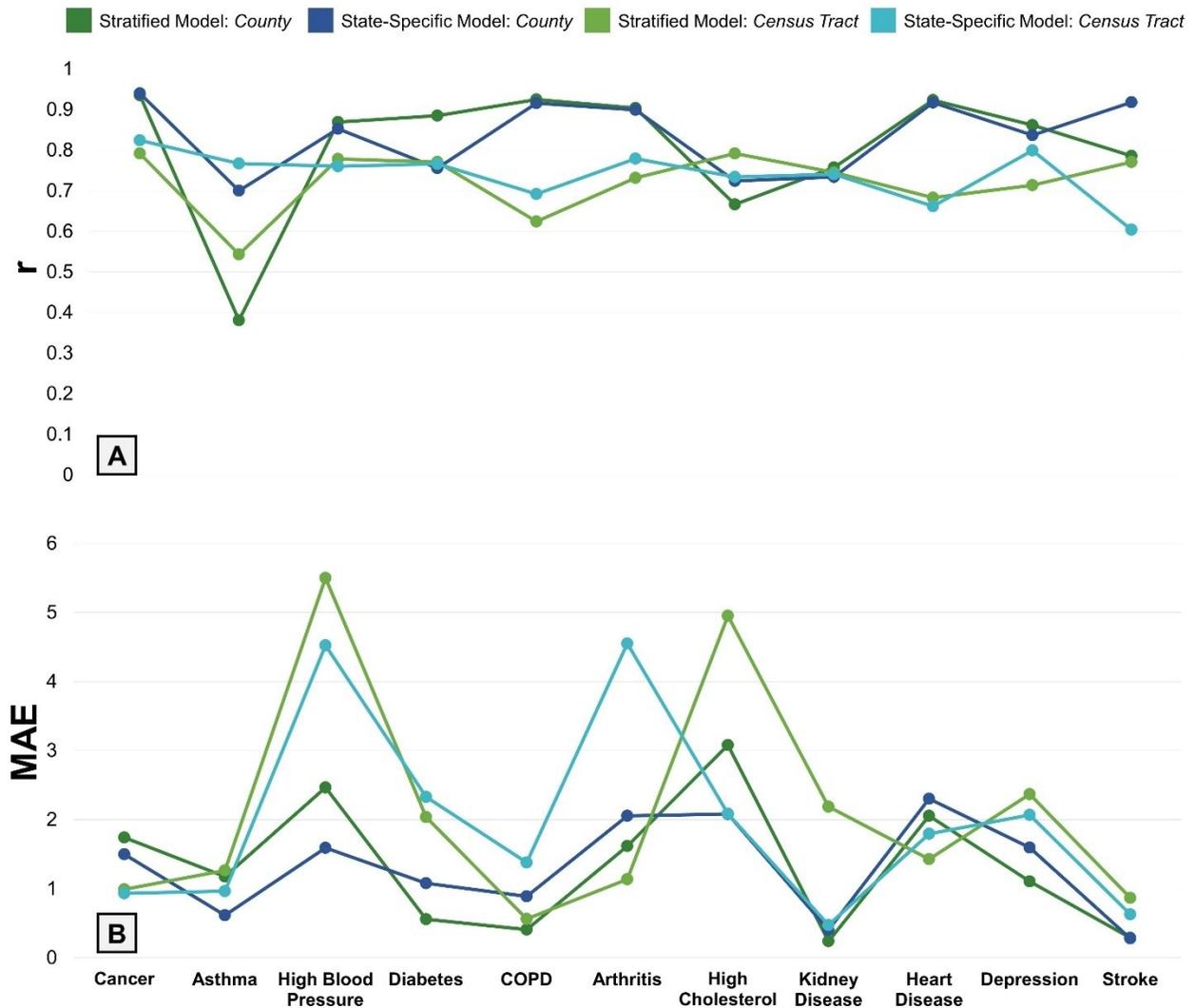

*Figure 7. Line charts evaluating both county and census tract level estimates from SHAPE models against CDC PLACES across health outcomes in **New York, 2021**, using the following metrics: A) Pearson's correlation coefficient (r), and B) Mean Absolute Error (MAE).*

### 3.2.1. Consistency with CDC PLACES

Overall, SHAPE demonstrated high consistency with CDC PLACES across both spatial scales, with only slight variations in agreement between models and states. For the county-level estimates, when averaged across the eleven health outcomes, SHAPE's Stratified Model showed slightly lower agreement in Florida ($\bar{r}$ = 0.743, average MAE = 1.650) than in New York ($\bar{r}$ = 0.809, average MAE = 1.341). The State-Specific Model performed marginally better than the Stratified Model in both states, with Florida ($\bar{r}$ = 0.837, average MAE = 1.228) and New York ($\bar{r}$ = 0.836, average MAE = 1.309), showing minimal differences between them.

At the census tract level, patterns in Pearson's r and MAE were generally consistent with those observed at the county level, although slightly lower r values and higher MAE reflect increased local variation across tracts. This greater variation arises because smaller geographic units contain fewer individuals, resulting in more fluctuation in estimates and slightly larger differences between SHAPE and CDC PLACES values. When averaged across the eleven health outcomes, the Stratified Model produced slightly higher agreement in Florida ($\bar{r}$ = 0.758, average MAE = 1.966) than in New York ($\bar{r}$ = 0.722, average MAE = 2.118). The State-Specific Model performed similarly to the Stratified

Model in both states, with Florida ($\bar{r}$ = 0.736, average MAE = 1.951) and New York ($\bar{r}$ = 0.739, average MAE = 2.080), again showing only minor differences between states.

*Pearson's Correlation.* Across both Florida (2019) and New York (2021), SHAPE's estimates demonstrated strong agreement ($r$ > 0.7) with CDC PLACES at the county level for most health outcomes (see dark blue and dark green lines in **Figure 6** and **Figure 7**). The only exception was asthma, with a negative correlation in Florida. Very high correlations ($r$ > 0.9) between SHAPE and CDC PLACES estimates were observed for high blood pressure, COPD, arthritis and stroke in Florida **(Figure 6A)**, and for COPD, cancer, high blood pressure and heart disease in New York **(Figure 7A)**. At the census tract level, similar patterns emerged, with high consistency between SHAPE and CDC PLACES estimates (see light blue and light green lines in **Figure 6** and **Figure 7**). Most health outcomes maintained strong correlations ($r$ > 0.6). Once again, the only exception was asthma, but this time with a positive relationship in Florida. The most notable differences in correlations between county- and census tract-level estimates were for asthma and kidney disease in Florida, and for COPD and kidney disease in New York. Overall, there were no major differences between the two states, indicating strong consistency between SHAPE and CDC PLACES across locations.

*Mean Absolute Error.* Patterns in MAE indicated close correspondence between SHAPE and CDC PLACES across outcomes in both states at the county level. SHAPE produced MAE values below 2.0 for most outcomes, with the exception of high cholesterol and heart disease in both Florida and New York, suggesting relatively small differences in the magnitude of estimates between the two sources. Significantly lower MAE values (≤ 1.0) were observed for several outcomes in both states, including asthma, diabetes, COPD, kidney disease, and stroke. At the census tract level, MAE values remained relatively low overall. However, MAE exceeded 2.0 for high blood pressure, arthritis, high cholesterol, and heart disease in Florida **(Figure 6B)**, and for high blood pressure, high cholesterol, and depression in New York **(Figure 7B)**. Notably, high blood pressure in New York showed a larger error at this scale, with an MAE greater than 4. Generally, MAE patterns were similar between county- and census tract-level estimates across most health outcomes, with the main differences observed for high blood pressure, arthritis, and COPD in Florida, and for high blood pressure, arthritis, high cholesterol, and kidney disease in New York.

### 3.2.2. Spatial Heterogeneity

**Figure 8** illustrates county-level estimates of heart disease prevalence in Florida (2019) and New York (2021) generated from SHAPE's Stratified Model (Panels A and C), and CDC PLACES (Panels B and D). Across both states, SHAPE captures the same spatial heterogeneity as CDC PLACES, but estimates slightly lower prevalence overall, explaining the higher r and MAE. In Florida, both models identify higher prevalence in northern counties. In New York, higher prevalence is also evident in counties such as Hamilton County (visible as the vertical rectangular area shaded in darker blue in the north-central region). Although visual differences between SHAPE and CDC PLACES may appear more pronounced, this is largely a result of the symbology: heart disease prevalence varies within a narrow range, and the use of 1% classification intervals amplifies small numerical differences between the two models.

**Figure 9** presents census tract–level estimates of arthritis prevalence in Florida (2019) and New York (2021) from SHAPE's Stratified Model (Panels A and C) and CDC PLACES (Panels B and D). Across both states, SHAPE and CDC PLACES produce similar spatial distributions, with only minor differences in magnitude. In Florida, both models identify a concentration of higher arthritis prevalence in central census tracts, with prevalence gradually declining toward the southern part of the state. In New York, both models similarly show higher prevalence clustered in centrally located tracts and along the southern border counties, while lower prevalence is observed in the southeastern region, including New York City and Long Island.

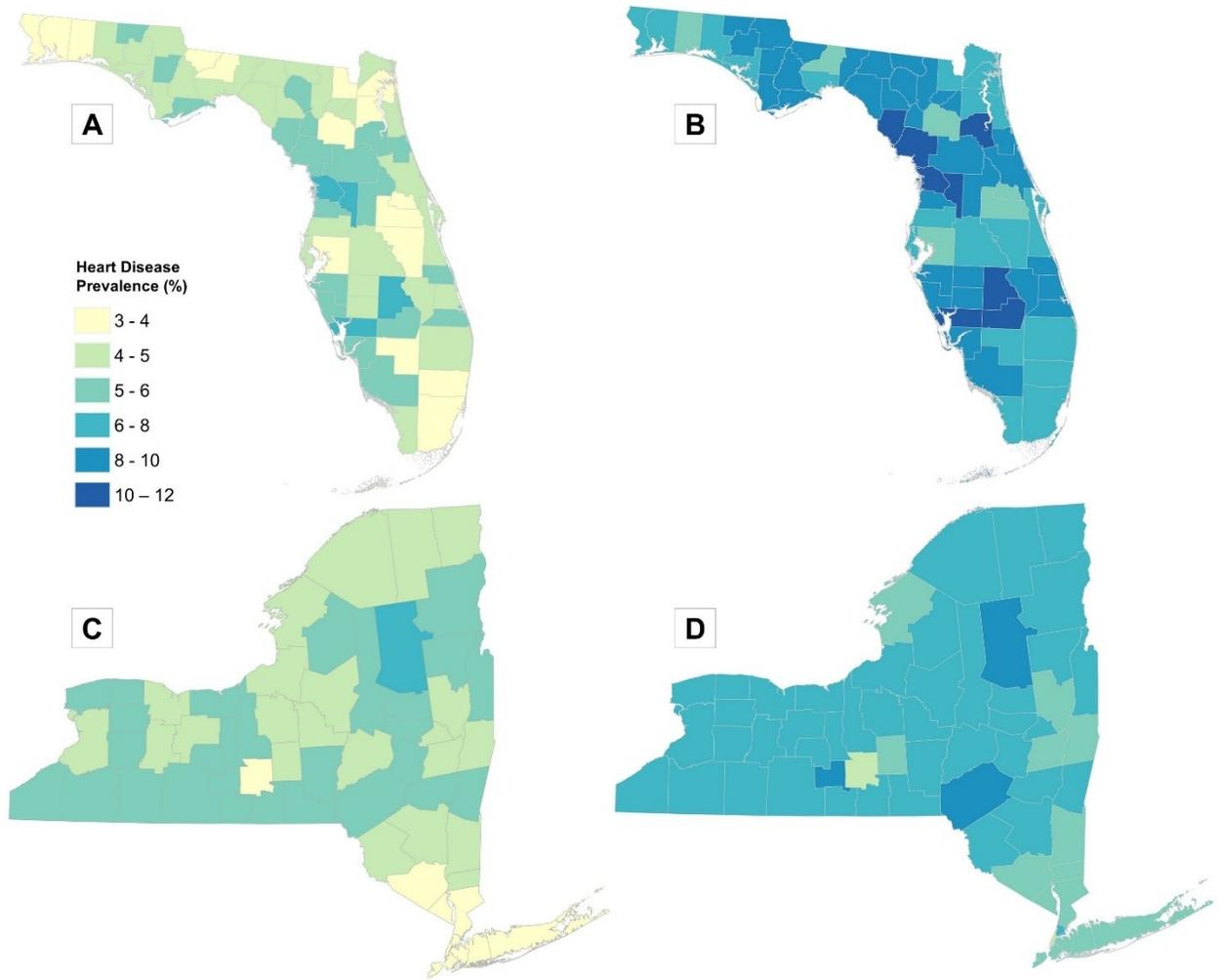

*Figure 8. County-level maps of heart disease prevalence estimates from the following sources: A) SHAPE's Stratified Model, Florida 2019; B) CDC PLACES estimates, Florida 2019; C) SHAPE's Stratified Model, New York 2021; D) CDC PLACES estimates, New York 2021.*

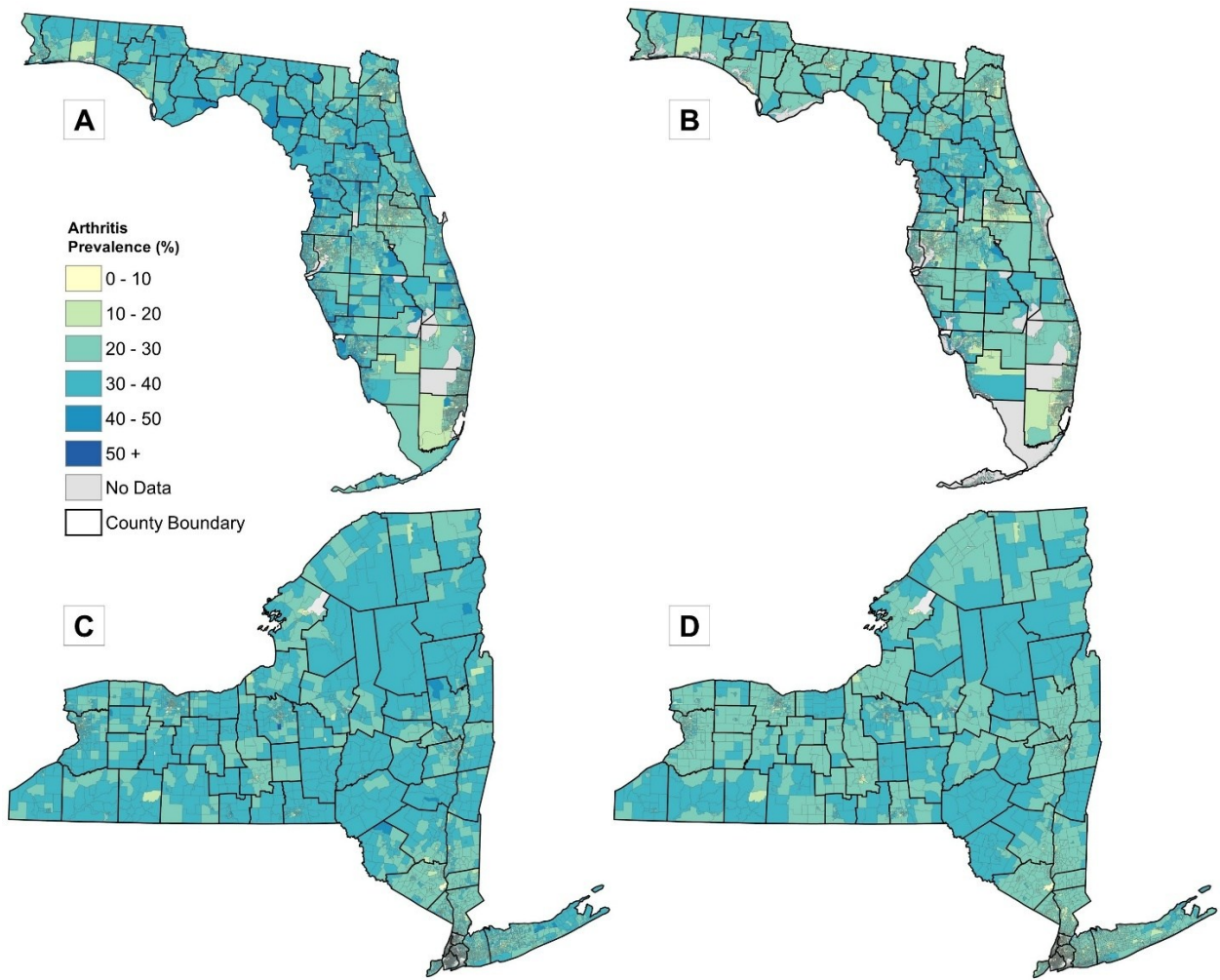

*Figure 9. Census tract-level maps of arthritis prevalence estimates from the following sources: A) SHAPE's Stratified Model, Florida 2019; B) CDC PLACES estimates, Florida 2019; C) SHAPE's Stratified Model, New York 2021; D) CDC PLACES estimates, New York 2021.*

### 3.3. Composite Scores and Model Ranking

While we focus here on the best performing model (SHAPE's State-Specific Model), a complete set of the calculated deviations from the evaluation metrics and the corresponding composite scores for all models are presented in the ***Appendix, Tables 13 and 14***. The composite scores range from 1.03 to –0.58 for the SHAPES State-Specific Model and 1.42 to –1.13 for CDC PLACES **(see Table 1 for model rankings).** Given that CDC PLACES estimates are used to inform public health agencies, we use the lowest *positive* composite score (0.8) for CDC PLACES as a threshold to identify SHAPE estimates that can be reliability used in public health policy settings. SHAPE's State-Specific Model achieved composite scores above 0.8 for heart disease, stroke, and diabetes. COPD, arthritis, and high blood pressure exceeded a composite score of 0.5 (the average across all models and outcomes), suggesting moderate reliability. Other measures, including cancer, smoking, high cholesterol, and depression, had lower positive composite scores ranging from 0.44 to 0.32, meaning that estimates for such outcomes should be used with caution. In contrast, SHAPES estimates for asthma, kidney disease, and obesity yielded negative scores, indicating limited reliability. CDC PLACES also yielded a negative score for kidney disease and heart disease since none of their estimates fall within the CIs of the direct estimates.

*Table 1. Composite scores, ordered greatest to least, from the ranking analysis summarizing model consistency with direct estimates across health behaviors and outcomes for SHAPE's State-Specific Model and CDC PLACES. An asterisk (\*) denotes a health risk behavior estimated in Level 1 of SHAPE's hierarchical IPF process (see Figure 1).*

| SHAPE's State-Specific Model | | CDC PLACES | |
|---|---|---|---|
| Health Measure Ranked | Composite Score | Health Measure Ranked | Composite Score |
| Heart Disease | 1.03 | Arthritis | 1.42 |
| Stroke | 0.94 | High Blood Pressure | 1.39 |
| Diabetes | 0.91 | Diabetes | 1.26 |
| COPD | 0.69 | Cancer | 1.19 |
| Arthritis | 0.65 | High Cholesterol | 1.10 |
| High Blood Pressure | 0.64 | Obesity* | 1.09 |
| Cancer | 0.44 | COPD | 0.98 |
| Smoking* | 0.42 | Stroke | 0.97 |
| High Cholesterol | 0.38 | Depression | 0.91 |
| Depression | 0.32 | Asthma | 0.88 |
| Asthma | -0.08 | Smoking* | 0.84 |
| Kidney Disease | -0.23 | Kidney Disease | -0.25 |
| Obesity* | -0.58 | Heart Disease | -1.13 |

## 4. Discussion

This study evaluated the performance of the SHAPE framework for producing SAEs of health behaviors and outcomes. SHAPE estimates showed moderate consistency with BRFSS direct estimates ($\bar{r} \sim 0.5$), similar to CDC PLACES ($\bar{r} \sim 0.6$). SHAPE estimates showed stronger consistency with CDC PLACES model-based estimates at the county level ($\bar{r} \sim 0.8$) and census tract level ($\bar{r} \sim 0.7$). In general, consistency varied somewhat by location, time, and the specific health outcome being modeled. In general, results from the composite scoring suggest that SHAPE could be used to generate estimates across the US for heart disease, stroke, and diabetes that could be used in policy settings as CDC PLACES is.

SHAPE is open, reproducible, and transparent, implemented entirely in R using publicly available data sources. Importantly, this addresses a growing and timely need for adaptable and sustainable SAE in public health. As national public health surveillance systems often experience changing priorities and resource limitations, there is an increasing need for transparent and scalable tools like SHAPE that can sustain local and state health monitoring capacity. By providing a flexible framework that can be implemented using publicly available data, SHAPE helps ensure that local health departments and researchers can continue rapidly generating reliable population health estimates even when access to federal model-based estimates or survey data is limited.

We evaluated SHAPE's robustness to different input data sources by applying the framework to both stratified and state-specific BRFSS samples, but also the CDC's nationally representative National Health Interview Survey (NHIS) (CDC, 2025b; see **Table 15 in the Appendix** for a subset of comparable results evaluating SHAPE's health risk and outcome estimates against direct estimates). BRFSS is a state-level telephone survey, whereas NHIS is a nationally representative, in-person household survey. We find that no matter the survey, SHAPE produced comparable SAEs, indicating that IPF is robust to differences in survey design, weighting, and sampling frame. Overall, these results highlight SHAPE's flexibility in terms of individual-level data requirements, enhancing its utility across contexts and applications, potentially beyond public health.

Existing published work evaluating CDC PLACES compares its estimates for five outcomes derived from the 2011 BRFSS (COPD, smoking, obesity, diabetes, and lack of health insurance) for a single state (Missouri) against county-level direct estimates (Zhang et al., 2015), with the model then applied to estimate additional health outcomes nationwide. Beyond providing an open framework for microsimulation-based SAE, our study offers further insights into the consistency of CDC PLACES estimates for multiple health outcomes and locations relative to direct survey data, extending the evaluation beyond its original scope.

While BRFSS direct estimates served as the primary point of comparison in the evaluation, it is important to recognize that these estimates are not ground truth data (i.e., fine-scale, directly observed health outcomes). To highlight why consistency with direct estimates does not necessarily imply greater accuracy, we present an example using county-level 2021 cancer prevalence data from the Surveillance, Epidemiology, and End Results (SEER) program for New York. Unlike BRFSS direct estimates, SEER data are based on cancer registry records and may therefore more closely approximate ground truth. We find that both SHAPE models had a substantially higher agreement with SEER cancer prevalence than with BRFSS direct estimates. For example, the Stratified Model resulted in r = 0.752 and MAE = 0.308, while CDC PLACES also showed strong performance with r = 0.783 and MAE = 1.416. In contrast, correlations between SHAPE and the BRFSS direct estimates for cancer were much weaker (~ 0.3; see Results section) and presented larger MAE values. The comparison with SEER indicates that both SHAPE and CDC PLACES perform better when evaluated against higher-quality reference data. Yet, true ground truth datasets are scarce in public health. This persistent challenge limits rigorous validation of SAEs, leading studies to omit validation efforts or to rely on existing modeled or survey-based estimates for comparison (Edwards et al., 2011).

There are many practical advantages to the SHAPE framework for public health research and applications. SHAPE generates individual-level synthetic populations that preserve joint distributions across sociodemographic variables, thereby reducing the ecological bias inherent in area-level modeling. This enables more detailed health policy analyses, such as examining disparities in health outcomes across certain racial groups, income brackets, and geographic areas. For example, as illustrated in Figure 10, SHAPE's Stratified Model produced individual-level data with joint attributes that allowed identification of Florida census tracts with higher concentrations of individuals who have both heart disease and no health insurance. Furthermore, because SHAPE produces individual-level microdata, the framework can be directly integrated with agent-based or spatial microsimulation models to explore disease dynamics, evaluate policy interventions, and conduct scenario analyses.

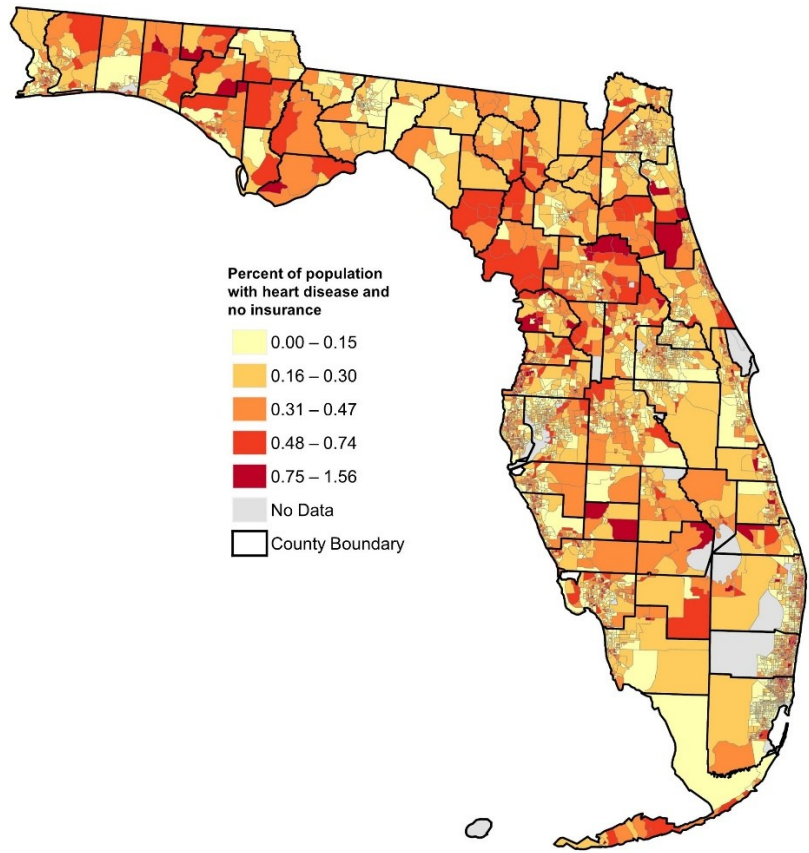

*Figure 10. Distribution of percentage of population with heart disease and no health insurance across Florida census tracts (2019). The population with such joint characteristics were generated with SHAPE's Stratified Model.*

Despite SHAPE's demonstrated utility, such microsimulation-based SAE approaches remain less widely applied in public health compared to more traditional methods due to overarching methodological challenges. One of the major limitation lies in the interpretability of their estimates for policy use since the typically used deterministic techniques like IPF do not produce measures of uncertainty (e.g., confidence intervals), which makes it difficult to quantify variability around model estimates (Tanton, 2017). However, some studies have begun exploring complementary techniques to address uncertainty in microsimulation-based SAE (e.g., Whitworth et al., 2017). Concerns about transparency and reproducibility also persist, as many existing microsimulation studies have relied on specialized software or lacked publicly available code. While the shift toward open-source tools such as R and Python has improved accessibility, reluctance to share code and data remains a barrier to broader adoption (Schofield et al., 2017; Smith et al., 2021).

Generally, most SAE approaches, including microsimulation and multilevel regression modeling, depend heavily on the quality and explanatory power of the predictor variables. Because these approaches often rely on sociodemographic predictors (e.g., income, education, race, age), their performance is tied to how strongly these variables explain variation in health outcomes. This limitation was evident in SHAPE's results: health outcomes such as high blood pressure and COPD that are more clearly linked to sociodemographic characteristics (as observed in regression models using the individual-level data - see ***Appendix, Table 2***) produced SAE with greater consistency. In contrast, outcomes like asthma and depression, which are likely more influenced by environmental or behavioral factors that could not be included in the IPF approach, produced weaker SAE. To partially address this, SHAPE incorporates a hierarchical structure in which health risk behaviors (e.g., smoking, obesity) are first simulated and then used as additional inputs for estimating health outcomes (see Figure 1). When we previously conducted sensitivity tests, models that integrated health risk behaviors achieved higher consistency metrics than those relying on sociodemographic variables alone, supporting the value of this multi-level approach.

Together, these limitations highlight several directions for advancing SHAPE and similar SAE frameworks. Incorporating additional behavioral and environmental determinants from publicly available sources may improve prediction of health outcomes that are not as strongly explained by demographics alone. Uncertainty quantification—one of the major limitations of microsimulation, as deterministic techniques like IPF do not capture variability—is an emerging area in which complementary methods are being developed for microsimulation-based SAEs. Incorporating such approaches into SHAPE, for example with bootstrapping or probabilistic reweighting, could improve interpretability for health policy applications. There is also an opportunity to assess potential bias and refine SHAPE's methodology through further calibration, such as sensitivity testing of different combinations of demographic predictors to optimize health outcome estimates or adjusting the stratified sampling procedures used for the input survey to better capture uncertainty across results. While this study uses IPF for its flexibility and well-established presence in the literature, future work could adapt SHAPE to incorporate alternative spatial microsimulation techniques such as CO or GREGWT and compare their performance. Additionally, since SHAPE produces agent-level outputs, it offers an opening to link SAE with dynamic simulation methods to explore population health dynamics under intervention scenarios. Beyond public health, SHAPE's scalable and flexible design could also be extended to generate SAES for other domains such as disaster management, marketing, and national security applications.

## 5. Conclusion

This study demonstrates the potential for spatial microsimulation as a scalable approach for SAE in public health. By generating individual-level datasets that preserve the joint distributions of characteristics such as sociodemographics and health outcomes, spatial microsimulation enables detailed analyses of health disparities at fine spatial scales and supports downstream modeling approaches, such as agent-based simulations. We introduced the Spatial Health and Population Estimator (SHAPE), which extends previous microsimulation studies by producing estimates for multiple health risk behaviors and outcomes across different locations, time periods, and spatial scales. Our evaluation shows that SHAPE generates health estimates moderately consistent with BRFSS direct estimates and strongly aligned with CDC PLACES, the gold standard for small-area estimation in public health, with particularly strong agreement at both county and census tract levels. SHAPE uses a two-stage hierarchical IPF workflow to produce SAEs of two health risk behaviors and eleven additional health outcomes for Florida (2019) and New York (2021). In the first stage, IPF aligns individual-level BRFSS survey data with aggregated sociodemographic data to estimate smoking and obesity prevalence. In the second stage, these estimates, together with sociodemographic variables, are used to model the eleven additional health outcomes. This work presents SHAPE as a transparent, reproducible, and adaptable tool capable of generating reliable multi-outcome health estimates across multiple contexts. Its flexibility and scalability address the growing need for robust SAE methods in public health, offering a timely resource for research, policy-making, and targeted interventions aimed at reducing health inequalities.

# Appendix for Evaluation of A Spatial Microsimulation Framework for Small-Area Estimation of Population Health Outcomes Using the Behavioral Risk Factor Surveillance System

## A. Descriptive Statistics: Individual-Level Surveys

Table 1. Percentage of respondents by sociodemographic characteristics, health risk behaviors, and health outcomes across the four input surveys used in the SHAPE framework and their corresponding original BRFSS source surveys.

| *Variable: Descriptor* | 2019: BRFSS (N=418,268) | 2019: Stratified Sampled (N=14,968) | 2019: Florida Sampled (N=10,473) | 2021: BRFSS (N=438,693) | 2021: Stratified Sampled (N=14,981) | 2021: New York Sampled (N=20,925) |
|---|---|---|---|---|---|---|
| *Gender: Male* | 47.317 | 47.334 | 46.510 | 48.151 | 48.148 | 48.344 |
| *Gender: Female* | 52.683 | 52.666 | 53.490 | 51.849 | 51.852 | 51.656 |
| *Race/Ethnicity: White* | 79.222 | 79.443 | 76.158 | 78.635 | 78.740 | 78.174 |
| *Race/Ethnicity: Black* | 7.537 | 7.449 | 9.186 | 7.261 | 7.256 | 7.039 |
| *Race/Ethnicity: Hispanic* | 6.682 | 6.648 | 9.205 | 7.010 | 6.969 | 8.225 |
| *Race/Ethnicity: Other* | 6.559 | 6.460 | 5.452 | 7.095 | 7.036 | 6.562 |
| *Age: 18-29* | 8.661 | 8.652 | 8.126 | 6.517 | 6.522 | 5.935 |
| *Age: 30-49* | 24.484 | 24.479 | 21.551 | 26.573 | 26.627 | 23.608 |
| *Age: 50-64* | 30.159 | 30.151 | 28.664 | 30.362 | 30.312 | 31.756 |
| *Age: 65+* | 36.696 | 36.718 | 41.660 | 36.548 | 36.540 | 38.700 |
| *Education: No Bachelor's Degree* | 41.623 | 41.542 | 34.049 | 45.031 | 44.924 | 42.891 |
| *Education: Bachelor's Degree or Higher* | 58.377 | 58.458 | 65.951 | 54.969 | 55.076 | 57.109 |
| *Income: Less than $25,000* | 22.632 | 22.588 | 29.848 | 14.591 | 14.619 | 16.812 |
| *Income: $25,000 - $49,999* | 23.647 | 23.657 | 27.089 | 25.418 | 25.406 | 26.934 |
| *Income: $50,000 - $74[1] / 99,999[2]* | 16.528 | 16.529 | 15.325 | 32.311 | 32.301 | 31.427 |
| *Income: Greater than $ 75[1] / 100,000[2]* | 37.193 | 37.226 | 27.738 | 27.680 | 27.675 | 24.827 |
| *Residency: Urban Area* | 69.119 | 69.161 | 73.952 | 69.588 | 69.642 | 71.345 |
| *Residency: Not Urban Area* | 30.881 | 30.839 | 26.048 | 30.412 | 30.358 | 28.655 |
| *Health Insurance: Insured* | 93.199 | 93.399 | 89.459 | 96.196 | 96.322 | 97.319 |
| *Health Insurance: Not Insured* | 6.801 | 6.601 | 10.541 | 3.804 | 3.678 | 2.681 |
| *Health Risk Behavior: Smoking* | 13.702 | 13.683 | 16.490 | 12.484 | 12.369 | 14.127 |
| *Health Risk Behavior: Obesity* | 33.627 | 33.071 | 32.130 | 35.890 | 35.638 | 35.149 |
| *Health Outcome: Arthritis* | 34.624 | 34.487 | 36.446 | 35.149 | 35.405 | 36.889 |
| *Health Outcome: Asthma* | 9.402 | 9.433 | 8.775 | 9.899 | 9.666 | 10.791 |
| *Health Outcome: Cancer* | 8.176 | 8.051 | 8.021 | 7.918 | 7.710 | 8.784 |
| *Health Outcome: High Cholesterol* | 37.237 | 36.966 | 38.652 | 40.055 | 39.624 | 40.898 |
| *Health Outcome: High Blood Pressure* | 41.672 | 41.422 | 44.830 | 42.026 | 41.613 | 41.840 |
| *Health Outcome: Kidney Disease* | 3.967 | 3.855 | 4.984 | 4.167 | 4.212 | 4.234 |
| *Health Outcome: COPD* | 8.435 | 8.665 | 12.461 | 8.018 | 7.843 | 9.190 |
| *Health Outcome: Kidney Disease* | 5.958 | 5.779 | 7.658 | 5.868 | 5.928 | 6.232 |
| *Health Outcome: Depression* | 19.548 | 19.568 | 19.641 | 20.411 | 20.746 | 20.846 |
| *Health Outcome: Diabetes* | 14.114 | 13.876 | 16.538 | 14.251 | 14.185 | 15.125 |
| *Health Outcome: Stroke* | 4.436 | 4.309 | 6.025 | 3.997 | 4.058 | 3.823 |

[1]*For 2019 BRFSS, stratified and Florida-sampled surveys;* [2]*For 2021 BRFSS, stratified and New York-sampled surveys*

B.  *Regressions Between Individual-Level Sociodemographics and Health Measures*

Table 2. Logistic regression coefficients and their statistical significance at the 95% confidence level (indicated by an asterisk *) for associations between sociodemographic predictor variables (used in the SHAPE framework) and health risk behaviors and outcomes across the four input surveys. All variables achieved VIF scores below 2.

| Variable ($R^2$) | Male | Black | Age 65+ | Bachelors or Higher | Income $100,000 + | Urban Residence | Health Insured | Smoking | Obesity |
|---|---|---|---|---|---|---|---|---|---|
| **Stratified Sampled Survey: 2021** | | | | | | | | | |
| Smoking (0.072) | 0.044 | -0.027* | -0.803* | -0.974* | -0.870* | -0.206* | -0.204 | NA | NA |
| Obesity (0.022) | -0.072* | 0.584* | -0.432* | -0.378* | -0.264* | -0.150* | 0.194* | NA | NA |
| Cancer | -0.271* | 0.094 | 1.105* | -0.153* | -0.046 | -0.044 | 0.473 | 0.067 | 0.077 |
| Asthma (0.039) | -0.764* | 0.149 | -0.254* | -0.041 | -0.236* | 0.052 | 0.135 | 0.171* | 0.622* |
| High Blood Pressure (0.111) | 0.399* | 0.582* | 1.322* | -0.262* | -0.193* | -0.076 | 0.335* | 0.285* | 0.989* |
| Diabetes (0.087) | 0.239* | 0.504* | 0.958* | -0.354* | -0.449* | -0.018 | 0.243 | 0.012 | 1.052* |
| COPD (0.104) | -0.172* | -0.191 | 0.828* | -0.571* | -0.803* | -0.016 | 0.319 | 1.429* | 0.476* |
| Arthritis (0.081) | -0.340* | -0.085 | 1.171* | -0.204* | -0.216* | -0.051 | 0.789* | 0.304* | 0.555* |
| High Cholesterol (0.045) | 0.139* | -0.053 | 0.971* | -0.040 | 0.015 | 0.062 | 0.442* | 0.178* | 0.426* |
| Kidney Disease (0.049) | 0.025 | 0.149 | 1.138* | -0.275* | -0.232* | -0.003 | 0.479 | -0.118* | 0.421* |
| Heart Disease (0.089) | 0.652* | -0.018 | 1.513* | -0.251* | -0.316* | -0.064 | 0.204 | 0.404* | 0.414* |
| Depression (0.062) | -0.768* | -0.706* | -0.546* | 0.077 | -0.539* | 0.180* | 0.727 | 0.727* | 0.455* |
| Stroke (0.065) | 0.120 | 0.656* | 1.075* | -0.452* | -0.721* | -0.070 | 0.564 | 0.545* | 0.013 |
| **New York Sampled Survey: 2021** | | | | | | | | | |
| Smoking (0.077) | 0.049 | -0.148 | -0.834* | -1.042* | -0.921* | -0.152* | -0.101 | NA | NA |
| Obesity (0.017) | -0.076* | 0.205* | -0.343* | -0.407* | -0.221* | -0.200* | 0.405* | NA | NA |
| Cancer | -0.235* | -0.218 | 1.129* | 0.014 | -0.227* | 0.030 | 0.700* | 0.045 | 0.190* |
| Asthma (0.037) | -0.665* | -0.134 | -0.352* | 0.050 | -0.280* | -0.068 | 0.367* | 0.199* | 0.612* |
| High Blood Pressure (0.101) | 0.457* | 0.494* | 1.227* | -0.227* | -0.228* | -0.090* | 0.651* | 0.015 | 0.912* |
| Diabetes (0.072) | 0.341* | 0.347* | 0.709* | -0.344* | -0.529* | -0.048 | 0.954* | 0.084 | 0.956* |
| COPD (0.111) | -0.096 | -0.519* | 0.873* | -0.689* | -0.656* | -0.146* | 0.708* | 1.419* | 0.354* |
| Arthritis (0.081) | -0.389* | -0.364* | 1.089* | -0.228* | -0.246* | -0.140* | 0.686* | 0.266* | 0.599* |
| High Cholesterol (0.034) | 0.152* | -0.168* | 0.780* | -0.020 | -0.057 | -0.013 | 0.613* | 0.138* | 0.381* |
| Kidney Disease (0.044) | 0.059 | 0.203 | 0.854* | -0.228* | -0.685* | -0.106 | 0.683* | 0.042 | 0.514* |
| Heart Disease (0.087) | 0.726* | -0.345* | 1.406* | -0.281* | -0.310* | 0.019 | 1.316* | 0.189* | 0.381* |
| Depression (0.056) | -0.662* | -0.578* | -0.468 | 0.023 | -0.544* | 0.065 | 0.398* | 0.723* | 0.456* |
| Stroke (0.052) | 0.240* | 0.353* | 0.997* | -0.322* | -0.697* | -0.142 | 0.712* | 0.483* | 0.310* |
| **Stratified Sampled Survey: 2019** | | | | | | | | | |
| Smoking (0.081) | -0.024 | 0.052 | -0.832* | -1.027* | -0.706* | -0.130* | -0.442* | NA | NA |
| Obesity (0.015) | 0.027 | 0.503* | -0.305* | -0.308* | -0.202* | -0.183* | 0.210* | NA | NA |
| Cancer (0.043) | -0.266* | -0.050 | 1.006* | -0.104 | -0.192* | 0.014 | 0.545* | 0.266* | -0.064 |
| Asthma (0.038) | -0.747* | -0.102 | -0.321* | -0.092 | -0.367* | 0.224* | 0.370* | 0.274* | 0.509* |
| High Blood Pressure (0.101) | 0.339* | 0.563* | 1.300* | -0.247* | -0.254* | -0.135* | 0.265* | 0.160* | 0.962* |
| Diabetes (0.088) | 0.312* | 0.467* | 0.901* | -0.358* | -0.409* | -0.026 | 0.221* | 0.034 | 1.101* |
| COPD (0.114) | -0.211* | -0.212 | 0.838* | -0.746* | -0.762* | -0.169* | 0.330* | 1.349* | 0.400* |
| Arthritis (0.087) | -0.431* | -0.073 | 1.128* | -0.289* | -0.256* | -0.144* | 0.475* | 0.283* | 0.600* |
| High Cholesterol (0.043) | 0.118* | -0.038 | 0.864* | -0.069 | -0.005 | -0.080* | 0.454* | 0.071 | 0.494* |
| Kidney Disease (0.046) | 0.103 | 0.198 | 0.977* | -0.228* | -0.482* | 0.199* | 0.464* | -0.039 | 0.470* |
| Heart Disease (0.081) | -0.590* | -0.448* | 1.429* | -0.260* | -0.014 | 0.161* | 0.189* | 0.389* | 0.436* |
| Depression (0.068) | -0.720* | -0.663* | -0.606* | 0.057 | -0.636* | 0.161* | 0.189* | 0.742* | 0.499* |
| Stroke (0.062) | 0.178* | 0.339* | 0.924* | -0.354* | -0.896* | -0.113 | 0.463* | 0.499* | 0.247* |
| **Florida Sampled Survey: 2019** | | | | | | | | | |
| **Smoking (0.084)** | 0.086 | -0.681* | -0.823* | -0.871* | -0.810* | -0.311* | -0.503* | NA | NA |
| **Obesity (0.015)** | -0.027 | 0.445* | -0.292* | -0.244* | -0.179* | -0.320* | 0.191* | NA | NA |
| **Cancer (0.039)** | -0.330* | 0.118 | 0.945* | -0.014 | -0.181 | -0.060 | 0.390* | 0.349* | 0.068 |
| **Asthma (0.038)** | -0.808* | 0.071 | -0.132 | -0.190* | -0.263* | -0.044 | 0.048 | 0.310* | 0.528* |
| **High Blood Pressure (0.094)** | 0.277* | 0.504* | 1.249* | -0.172* | -0.233* | -0.236* | 0.181* | 0.103 | 0.776* |
| **Diabetes (0.093)** | 0.314* | 0.610* | 0.940* | -0.256* | -0.510* | -0.209* | 0.513* | -0.265* | 0.977* |
| **COPD (0.011)** | -0.240* | -0.567* | 0.788* | -0.643* | -0.704* | -0.241* | 0.635* | NA | NA |
| **Arthritis (0.080)** | -0.469* | -0.208* | 1.049* | -0.179* | -0.243* | -0.094 | 0.476* | NA | NA |
| **High Cholesterol (0.045)** | -0.088* | -0.139 | 0.865* | -0.089 | -0.092 | -0.083 | 0.337* | 0.154* | 0.457* |
| **Kidney Disease (0.051)** | -0.019 | 0.540* | 0.992* | -0.183 | -0.406* | -0.255* | 0.589* | -0.082 | 0.331* |
| **Heart Disease (0.074)** | 0.349* | -0.327* | 1.297* | -0.316* | -0.336* | -0.190* | 0.709* | 0.149 | 0.242* |
| **Depression (0.061)** | -0.644* | -0.805* | -0.424* | -0.188* | -0.675* | -0.037 | 0.059 | 0.624* | 0.369* |
| **Stroke (0.058)** | 0.118 | 0.314* | 1.108* | -0.369* | -0.634* | -0.010 | 0.347 | 0.370* | 0.077 |

*Evaluation of A Spatial Microsimulation Framework for Small-Area Estimation of Population Health Outcomes Using the Behavioral Risk Factor Surveillance System*### C. Descriptive Statistics: Spatial Aggregated Sociodemographic Data

*Table 3. Descriptive statistics for spatially aggregated county-level sociodemographic variables used as input data for the SHAPE models in Florida (2019) and New York (2021).*

| | **New York: 2021 (N=62)** | | | | **Florida: 2019 (N = 67)** | | | |
|---|---|---|---|---|---|---|---|---|
| *Variable: Descriptor* | **Mean** | **Min** | **Max** | **Std.Dev.** | **Mean** | **Min** | **Max** | **Std.Dev.** |
| *Gender: Male* | 49.757 | 46.200 | 55.274 | 1.746 | 51.533 | 46.707 | 69.640 | 4.468 |
| *Gender: Female* | 50.243 | 44.726 | 53.800 | 1.746 | 48.467 | 30.360 | 53.293 | 4.468 |
| *Race/Ethnicity: White* | 81.336 | 9.948 | 95.768 | 17.082 | 68.352 | 12.368 | 88.262 | 15.127 |
| *Race/Ethnicity: Black* | 6.130 | 0.688 | 34.427 | 6.420 | 13.789 | 2.217 | 53.396 | 9.143 |
| *Race/Ethnicity: Hispanic* | 7.325 | 1.338 | 54.586 | 8.818 | 13.713 | 2.964 | 70.314 | 12.854 |
| *Race/Ethnicity: Other* | 5.210 | 1.038 | 29.677 | 4.429 | 4.147 | 1.210 | 8.966 | 1.768 |
| *Age: 18-29* | 19.841 | 11.333 | 38.900 | 4.202 | 17.361 | 5.917 | 35.852 | 4.695 |
| *Age: 30-49* | 29.580 | 20.486 | 37.808 | 2.952 | 30.251 | 13.190 | 39.964 | 5.075 |
| *Age: 50-64* | 27.193 | 19.302 | 31.819 | 2.267 | 25.450 | 18.817 | 29.241 | 2.191 |
| *Age: 65+* | 23.386 | 17.051 | 36.362 | 3.175 | 26.938 | 15.422 | 62.076 | 8.316 |
| *Education: No Bachelor's Degree* | 87.531 | 71.943 | 93.691 | 4.523 | 89.385 | 78.718 | 96.620 | 4.765 |
| *Education: Bachelor's Degree or Higher* | 12.469 | 6.309 | 28.057 | 4.523 | 10.615 | 3.380 | 21.282 | 4.765 |
| *Income: Less than $25,000* | 18.168 | 8.609 | 32.881 | 4.078 | 20.918 | 10.458 | 33.775 | 5.897 |
| *Income: $25,000 - $49,999* | 20.402 | 9.870 | 26.153 | 3.843 | 24.451 | 13.801 | 34.363 | 4.334 |
| *Income: $50,000 - $74[1] / 99,999[2]* | 30.686 | 20.065 | 36.506 | 3.759 | 18.670 | 11.080 | 25.289 | 2.114 |
| *Income: Greater than $ 75[1] / 100,000[2]* | 30.743 | 19.243 | 60.877 | 9.802 | 35.960 | 18.684 | 57.819 | 9.271 |
| *Residency: Urban Area* | 51.964 | 0.000 | 100.000 | 30.481 | 58.237 | 0.000 | 99.992 | 38.747 |
| *Residency: Not Urban Area* | 48.036 | 0.000 | 100.000 | 30.481 | 41.763 | 0.008 | 100.000 | 38.747 |
| *Health Insurance: Insured* | 95.250 | 82.698 | 97.300 | 2.084 | 87.242 | 76.201 | 94.300 | 3.363 |
| *Health Insurance: Not Insured* | 4.750 | 2.700 | 17.302 | 2.084 | 12.758 | 5.700 | 23.799 | 3.363 |

[1]*For 2019 spatial aggregate data;* [2]*For 2021 spatial aggregate data*

*Table 4. Descriptive statistics for spatially aggregated census tract-level sociodemographic variables used as input data for the SHAPE models in Florida (2019) and New York (2021).*

| | **New York: 2021 (N=5,264)** | | | | **Florida: 2019 (N = 5,070)** | | | |
|---|---|---|---|---|---|---|---|---|
| *Variable: Descriptor* | **Mean** | **Min** | **Max** | **Std.Dev.** | **Mean** | **Min** | **Max** | **Std.Dev.** |
| *Gender: Male* | 47.891 | 0.000 | 100.000 | 5.676 | 48.376 | 21.279 | 100.000 | 5.920 |
| *Gender: Female* | 52.109 | 0.000 | 100.000 | 5.676 | 51.624 | 0.000 | 78.721 | 5.920 |
| *Race/Ethnicity: White* | 57.354 | 0.000 | 100.000 | 33.927 | 57.573 | 0.000 | 100.000 | 29.450 |
| *Race/Ethnicity: Black* | 15.914 | 0.000 | 100.000 | 23.640 | 14.451 | 0.000 | 99.471 | 19.391 |
| *Race/Ethnicity: Hispanic* | 16.321 | 0.000 | 95.448 | 18.529 | 23.699 | 0.000 | 100.000 | 24.647 |
| *Race/Ethnicity: Other* | 10.412 | 0.000 | 100.000 | 14.371 | 4.277 | 0.000 | 47.881 | 4.505 |
| *Age: 18-29* | 20.693 | 0.000 | 100.000 | 9.839 | 17.934 | 0.000 | 100.000 | 10.394 |
| *Age: 30-49* | 32.702 | 0.000 | 100.000 | 8.574 | 30.164 | 0.000 | 100.000 | 10.970 |
| *Age: 50-64* | 25.519 | 0.000 | 100.000 | 6.904 | 25.118 | 0.000 | 59.615 | 6.780 |
| *Age: 65+* | 21.086 | 0.000 | 100.000 | 8.670 | 26.783 | 0.000 | 96.289 | 15.601 |
| *Education: No Bachelor's Degree* | 84.494 | 0.000 | 100.000 | 9.721 | 86.257 | 0.000 | 100.000 | 8.910 |
| *Education: Bachelor's Degree or Higher* | 15.506 | 0.000 | 100.000 | 9.721 | 13.743 | 0.000 | 100.000 | 8.910 |
| *Income: Less than $25,000* | 19.098 | 0.000 | 100.000 | 13.281 | 20.354 | 0.000 | 100.000 | 12.061 |
| *Income: $25,000 - $49,999* | 18.341 | 0.000 | 100.000 | 8.784 | 23.877 | 0.000 | 100.000 | 9.560 |
| *Income: $50,000 - $99,999* | 27.465 | 0.000 | 100.000 | 9.436 | 30.379 | 0.000 | 100.000 | 9.258 |
| *Income: Greater than $100,000* | 35.096 | 0.000 | 100.000 | 18.897 | 25.390 | 0.000 | 100.000 | 16.294 |
| *Residency: Urban Area* | 85.276 | 0.000 | 100.000 | 32.880 | 90.369 | 0.000 | 100.000 | 25.888 |
| *Residency: Not Urban Area* | 14.724 | 0.000 | 100.000 | 32.880 | 9.631 | 0.000 | 100.000 | 25.888 |
| *Health Insurance: Insured* | 94.587 | 52.179 | 100.000 | 4.570 | 87.405 | 28.057 | 100.000 | 7.830 |
| *Health Insurance: Not Insured* | 5.413 | 0.000 | 47.821 | 4.570 | 12.595 | 0.000 | 71.943 | 7.830 |

[1]*For 2019 spatial aggregate data;* [2]*For 2021 spatial aggregate data*

### D. Descriptive Statistics: Small Area Estimates of Health Measures

*Table 5. Descriptive statistics for **Florida 2019 health outcomes** from direct, SHAPE and CDC PLACES estimates.*

| | Cancer | Asthma | Hyper-tension | Diabetes | COPD | Arthritis | Kidney Disease | Heart Disease | Depress-Ion | Stroke | High Cholesterol |
|---|---|---|---|---|---|---|---|---|---|---|---|
| **Direct Estimates** | | | | | | | | | | | |
| **Mean** | 8.284 | 8.478 | 38.121 | 13.370 | 9.951 | 28.970 | 3.761 | 5.649 | 17.776 | 4.521 | 32.324 |
| **Std. Dev.** | 2.253 | 2.222 | 5.051 | 3.090 | 3.010 | 5.309 | 1.161 | 1.482 | 3.258 | 1.287 | 3.825 |
| **Min** | 4.200 | 4.600 | 25.300 | 6.400 | 4.000 | 17.800 | 1.700 | 2.500 | 10.300 | 1.200 | 23.600 |
| **Max** | 16.100 | 18.200 | 47.000 | 20.800 | 16.800 | 40.200 | 7.700 | 9.000 | 24.700 | 7.000 | 43.700 |
| **SHAPE: Stratified Model** | | | | | | | | | | | |
| **Mean** | 6.255 | 9.068 | 38.240 | 12.738 | 7.932 | 29.979 | 3.204 | 4.688 | 18.797 | 3.667 | 32.324 |
| **Std. Dev.** | 1.141 | 0.754 | 4.028 | 1.362 | 1.382 | 3.854 | 0.429 | 0.899 | 1.247 | 0.698 | 2.775 |
| **Min** | 4.184 | 6.841 | 30.340 | 10.059 | 4.841 | 21.868 | 2.580 | 3.023 | 15.901 | 2.214 | 25.660 |
| **Max** | 9.954 | 10.485 | 48.405 | 16.212 | 11.040 | 40.543 | 4.552 | 7.849 | 21.046 | 4.982 | 41.985 |
| **SHAPE: State-Specific Model** | | | | | | | | | | | |
| **Mean** | 5.890 | 8.253 | 37.976 | 13.621 | 9.840 | 28.788 | 4.039 | 5.408 | 18.678 | 4.425 | 32.063 |
| **Std. Dev.** | 0.940 | 0.571 | 4.472 | 2.073 | 1.558 | 3.461 | 0.697 | 1.076 | 1.548 | 0.799 | 3.064 |
| **Min** | 4.176 | 6.556 | 29.062 | 10.109 | 6.791 | 22.160 | 2.975 | 3.478 | 15.286 | 2.941 | 23.904 |
| **Max** | 9.249 | 10.077 | 48.640 | 18.846 | 13.272 | 39.464 | 6.184 | 9.227 | 22.255 | 6.547 | 41.684 |
| **CDC PLACES** | | | | | | | | | | | |
| **Mean** | 8.096 | 8.737 | 37.306 | 13.787 | 10.012 | 27.881 | 3.731 | 8.296 | 19.418 | 4.470 | 34.573 |
| **Std. Dev.** | 1.562 | 0.685 | 4.277 | 1.943 | 2.055 | 4.112 | 0.496 | 1.591 | 1.835 | 0.792 | 3.027 |
| **Min** | 5.400 | 7.500 | 27.700 | 9.500 | 6.000 | 19.300 | 2.800 | 5.200 | 15.400 | 3.000 | 26.900 |
| **Max** | 13.600 | 10.300 | 48.300 | 17.700 | 14.200 | 40.800 | 4.800 | 12.000 | 24.000 | 5.900 | 42.800 |

*Table 6. Descriptive statistics for **New York 2021 health outcomes** from direct, SHAPE and CDC PLACES estimates.*

|  | Cancer | Asthma | Hyper-tension | Diabetes | COPD | Arthritis | Kidney Disease | Heart Disease | Depress-Ion | Stroke | High Cholesterol |
|---|---|---|---|---|---|---|---|---|---|---|---|
| **Direct Estimates** | | | | | | | | | | | |
| Mean | 7.819 | 11.085 | 33.245 | 11.508 | 7.509 | 28.920 | 7.509 | 4.216 | 20.830 | 2.862 | 35.967 |
| Std. Dev. | 1.927 | 2.694 | 4.307 | 2.354 | 2.372 | 5.160 | 2.372 | 1.117 | 4.506 | 1.108 | 3.451 |
| Min | 3.287 | 4.721 | 20.300 | 6.200 | 2.541 | 18.517 | 2.541 | 1.946 | 8.598 | 1.120 | 24.817 |
| Max | 17.333 | 16.926 | 45.900 | 16.800 | 12.182 | 40.131 | 12.182 | 7.658 | 30.129 | 7.198 | 42.892 |
| **SHAPE: Stratified Model** | | | | | | | | | | | |
| Mean | 6.011 | 10.032 | 36.230 | 11.971 | 8.034 | 29.851 | 3.411 | 4.789 | 22.757 | 3.533 | 33.045 |
| Std. Dev. | 0.561 | 0.440 | 2.421 | 0.973 | 1.280 | 2.685 | 0.391 | 0.693 | 2.034 | 0.370 | 1.558 |
| Min | 4.341 | 8.868 | 28.953 | 9.238 | 5.369 | 23.554 | 2.639 | 3.277 | 17.854 | 2.551 | 26.961 |
| Max | 7.412 | 11.310 | 42.531 | 14.784 | 10.005 | 35.804 | 4.173 | 6.477 | 26.144 | 4.215 | 37.104 |
| **SHAPE: State-Specific Model** | | | | | | | | | | | |
| Mean | 6.252 | 11.619 | 35.094 | 12.623 | 8.864 | 30.824 | 3.554 | 4.544 | 23.637 | 2.998 | 34.171 |
| Std. Dev. | 0.594 | 0.718 | 2.366 | 1.173 | 1.667 | 3.127 | 0.531 | 0.628 | 2.351 | 0.353 | 1.628 |
| Min | 4.586 | 9.697 | 28.242 | 9.992 | 5.492 | 22.338 | 2.352 | 3.091 | 17.625 | 2.285 | 29.096 |
| Max | 7.934 | 12.683 | 42.305 | 15.140 | 11.458 | 37.031 | 4.706 | 6.388 | 26.939 | 3.978 | 38.668 |
| **CDC PLACES** | | | | | | | | | | | |
| Mean | 7.752 | 11.116 | 33.818 | 11.752 | 8.060 | 28.995 | 3.294 | 6.845 | 22.474 | 3.281 | 36.079 |
| Std. Dev. | 0.853 | 0.676 | 2.860 | 1.366 | 1.466 | 3.835 | 0.290 | 0.923 | 2.753 | 0.386 | 2.120 |
| Min | 5.200 | 9.100 | 24.100 | 8.200 | 4.800 | 19.800 | 2.400 | 4.700 | 14.800 | 2.300 | 28.100 |
| Max | 10.300 | 12.500 | 39.400 | 16.100 | 11.900 | 36.400 | 4.300 | 9.800 | 26.700 | 4.600 | 41.000 |

### E. Evaluation of Level 1: Health Risk Behavior Results
#### i. SHAPE and CDC PLACES vs. Direct Estimates

Health risk behavior estimates for smoking and obesity generated by the SHAPE framework, along with CDC PLACES estimates, are compared with direct estimates from the corresponding state and year, specifically Florida 2019 and New York 2021. Because SHAPE estimates were generated using both the stratified sampled survey and the state-specific sampled survey for each state, results are presented for both approaches. Evaluation metrics for overall accuracy are summarized in **Table 7** using Pearson's r coefficient, the coefficient ($R^2$) and Mean Absolute Error (MAE).

*Table 7. Evaluation metrics comparing CDC PLACES estimates and two SHAPE estimates (one from a model using the stratified sampled survey and the other from a model using the state-specific sampled survey) against direct survey estimates at the county level in Florida (2019) and in New York (2021).*

| | Stratified Model | | | | State-Specific Model | | | | CDC PLACES | | | |
|---|---|---|---|---|---|---|---|---|---|---|---|---|
| **SAE Outcome** | r | $R^2$ | MAE | % in CI | r | $R^2$ | MAE | % in CI | r | $R^2$ | MAE | % in CI |
| **Florida - 2019** | | | | | | | | | | | | |
| *Smoking* | 0.678 | 0.460 | 3.358 | 55.73 | 0.731 | 0.534 | 2.785 | 61.19 | 0.814 | 0.663 | 2.393 | 71.64 |
| *Obesity* | 0.767 | 0.589 | 4.665 | 55.22 | 0.752 | 0.566 | 4.133 | 58.21 | 0.861 | 0.741 | 2.555 | 67.16 |
| **New York - 2021** | | | | | | | | | | | | |
| *Smoking* | 0.720 | 0.519 | 2.926 | 79.03 | 0.726 | 0.526 | 2.907 | 64.52 | 0.812 | 0.659 | 2.615 | 66.13 |
| *Obesity* | 0.498 | 0.248 | 5.946 | 62.90 | 0.550 | 0.303 | 5.699 | 62.90 | 0.801 | 0.642 | 2.665 | 96.75 |

Correlations are relatively moderate for smoking in Florida (r = 0.678, 0.731; R² = 0.460, 0.534; MAE = 3.358, 2.785), but in contrast to New York, metrics are stronger for obesity (r = 0.767, 0.752; R² = 0.589, 0.566; MAE = 4.665, 4.133). CDC PLACES estimates again show the highest correlations for both outcomes (smoking: r = 0.814, R² = 0.663, MAE = 2.393; obesity: r = 0.861, R² = 0.741, MAE = 2.555). For smoking, 55.73% (n=36) of counties in the stratified SHAPE model and 61.19% (n=41) in the state-specific model fell within the BRFSS direct estimates' 95% confidence intervals, compared to 71.64% (n=48) for CDC PLACES. Coverage rates were slightly lower for obesity, with 55.22% (n=37) of counties within the intervals for the stratified SHAPE model and 58.21% (n=39) for the state-specific model, while CDC PLACES resulted in a slightly higher agreement at 67.16% (n=45).

For New York, both SHAPE approaches also show moderate correlation with smoking (r = 0.720, 0.726; R² = 0.519, 0.526; MAE ≈ 2.926, 2.907 for stratified and state-specific, respectively), while correlations for obesity are lower (r = 0.498, 0.550; R² = 0.248, 0.303; MAE = 5.946, 5.699). CDC PLACES achieves higher correlations for both outcomes (smoking: r = 0.812, R² = 0.659, MAE = 2.615; obesity: r = 0.801, R² = 0.642, MAE = 2.665) with values closer to SHAPE for smoking but a larger difference for obesity. In terms of coverage, 79.03% (n = 49), 64.52% (n = 40), and 66.13% (n = 41) of county-level smoking estimates from the stratified, state-specific, and CDC PLACES approaches, respectively, fell within the 95% confidence intervals of the BRFSS direct estimates. For obesity, the proportions were lower for SHAPE's stratified and state-specific models (62.90%, n = 39 for both) but substantially higher for CDC PLACES (96.75%, n = 60).

#### ii. SHAPE vs. CDC PLACES Estimates

Additionally, health risk behavior estimates for smoking and obesity generated from the SHAPE framework using both the stratified and state-specific sampled surveys are evaluated CDC PLACES estimates for Florida (2019) and New York (2021) at both the county and census tract levels (**Table 8**).

Table 8. Evaluation metrics comparing two SHAPE estimates (one from a model using the stratified sampled survey and the other from a model using the state-specific sampled survey) against CDC PLACES estimates at the county and census tract level in Florida (2019) and New York (2021).

|  | **Stratified Model** | | | | | | **State-Specific Model** | | | | | |
|---|---|---|---|---|---|---|---|---|---|---|---|---|
|  | **County** | | | **Census Tract** | | | **County** | | | **Census Tract** | | |
| **SAE Outcome** | r | $R^2$ | MAE | r | $R^2$ | MAE | r | $R^2$ | MAE | r | $R^2$ | MAE |
| **Florida - 2019** | | | | | | | | | | | | |
| *Smoking* | 0.924 | 0.854 | 2.997 | 0.869 | 0.755 | 3.245 | 0.930 | 0.865 | 1.908 | 0.599 | 0.358 | 3.821 |
| *Obesity* | 0.745 | 0.555 | 3.908 | 0.733 | 0.537 | 5.036 | 0.755 | 0.569 | 3.472 | 0.653 | 0.427 | 4.009 |
| **New York - 2021** | | | | | | | | | | | | |
| *Smoking* | 0.924 | 0.854 | 2.997 | 0.869 | 0.755 | 3.245 | 0.930 | 0.865 | 1.908 | 0.599 | 0.358 | 3.821 |
| *Obesity* | 0.745 | 0.555 | 3.908 | 0.733 | 0.537 | 5.036 | 0.755 | 0.569 | 3.472 | 0.653 | 0.427 | 4.009 |

In Florida, SHAPE estimates align closely with CDC PLACES at the county level. Smoking correlations are slightly higher than in New York (r = 0.924, 0.930; R² = 0.854, 0.865), with lower MAE for the state-specific survey (1.908 vs. 2.997). Obesity correlations are lower (r = 0.745, 0.755; R² = 0.555, 0.569), again with slightly better performance for the state-specific survey (MAE = 3.472 vs. 3.908). At the census tract level, correlations decrease for both outcomes, and MAE values increase, consistent with the pattern observed in New York. For smoking, correlations and variance remain relatively stable for the stratified survey (r = 0.869, R² = 0.755) but decline more significantly for the state-specific survey (r = 0.599, R² = 0.583), while MAE remains similar between the two approaches (3.245 vs. 3.821). Obesity estimates are also comparable between county- and tract-level analyses for both survey approaches (stratified: r = 0.733, R² = 0.537, MAE = 5.036; state-specific: r = 0.653, R² = 0.427, MAE = 4.009), with slightly higher MAE values, particularly for the stratified survey.

SHAPE estimates also show strong agreement with CDC PLACES at the county level in New York using both the stratified and state-specific sampled surveys. For smoking, correlations are high (r = 0.882, 0.895; R² = 0.778, 0.801), with lower MAE for the state-specific survey (1.062 vs. 1.812). Obesity estimates are weaker (r = 0.707, 0.733; R² = 0.500, 0.537), though MAE is again reduced with the state-specific survey (3.786 vs. 4.259). At the census tract level, correlations decrease slightly and MAE values increase for both outcomes, reflecting greater variability at finer spatial scales (smoking: r = 0.768, 0.672; R² = 0.590, 0.452; MAE = 3.122, 3.084; obesity: r = 0.699, 0.731; R² = 0.488, 0.534; MAE = 7.343, 5.528). The stratified survey better captures variability in smoking, whereas the state-specific survey slightly better captures variability in obesity, though with a notably higher MAE.

*Evaluation of A Spatial Microsimulation Framework for Small-Area Estimation of Population Health Outcomes Using the Behavioral Risk Factor Surveillance System*### F. Evaluating SHAPE and CDC PLACES Estimates Against Direct Estimates from BRFSS

Table 9. Evaluation metrics comparing SHAPE and CDC PLACES estimates against direct estimates from state-specific BRFSS surveys. * Indicates that the estimates were generated without health risk behaviors (smoking and obesity) as predictor variables.

| SAE Outcome | Stratified Model | | | State-Specific Model | | | CDC PLACES | | |
| --- | --- | --- | --- | --- | --- | --- | --- | --- | --- |
| | r | $R^2$ | MAE | r | $R^2$ | MAE | r | $R^2$ | MAE |
| **New York (2021)** | | | | | | | | | |
| Smoking* | 0.720 | 0.519 | 2.926 | 0.726 | 0.526 | 2.907 | 0.812 | 0.659 | 2.615 |
| Obesity* | 0.498 | 0.248 | 5.946 | 0.550 | 0.303 | 5.699 | 0.801 | 0.642 | 2.665 |
| Cancer | 0.319 | 0.102 | 1.963 | 0.331 | 0.109 | 1.796 | 0.341 | 0.116 | 1.147 |
| Asthma | 0.115 | 0.013 | 2.421 | 0.319 | 0.102 | 1.968 | 0.578 | 0.334 | 1.822 |
| High Blood Pressure | 0.643 | 0.414 | 3.688 | 0.624 | 0.389 | 3.009 | 0.869 | 0.755 | 1.853 |
| Diabetes | 0.565 | 0.320 | 1.680 | 0.594 | 0.353 | 1.850 | 0.692 | 0.479 | 1.408 |
| COPD | 0.666 | 0.443 | 1.562 | 0.612 | 0.374 | 1.848 | 0.687 | 0.472 | 1.501 |
| Arthritis | 0.748 | 0.559 | 3.017 | 0.746 | 0.557 | 3.170 | 0.889 | 0.790 | 1.916 |
| High Cholesterol | 0.318 | 0.101 | 3.699 | 0.352 | 0.124 | 3.127 | 0.663 | 0.439 | 1.933 |
| Kidney Disease | 0.610 | 0.372 | 4.105 | 0.534 | 0.285 | 3.959 | 0.303 | 0.092 | 4.230 |
| Heart Disease | 0.346 | 0.120 | 1.042 | 0.348 | 0.121 | 0.923 | 0.406 | 0.165 | 2.638 |
| Depression | 0.754 | 0.569 | 3.020 | 0.718 | 0.516 | 3.534 | 0.837 | 0.700 | 2.457 |
| Stroke | 0.170 | 0.029 | 1.055 | 0.325 | 0.106 | 0.845 | 0.322 | 0.103 | 0.884 |
| **Florida (2019)** | | | | | | | | | |
| Smoking* | 0.678 | 0.460 | 3.358 | 0.731 | 0.534 | 2.785 | 0.814 | 0.663 | 2.393 |
| Obesity* | 0.767 | 0.589 | 4.665 | 0.752 | 0.566 | 4.133 | 0.861 | 0.741 | 2.555 |
| Cancer | 0.696 | 0.484 | 2.132 | 0.685 | 0.469 | 2.451 | 0.749 | 0.561 | 1.125 |
| Asthma | -0.190 | 0.036 | 1.864 | -0.110 | 0.012 | 1.721 | 0.453 | 0.205 | 1.508 |
| High Blood Pressure | 0.764 | 0.584 | 2.664 | 0.765 | 0.586 | 2.676 | 0.844 | 0.713 | 2.327 |
| Diabetes | 0.559 | 0.312 | 2.184 | 0.632 | 0.399 | 1.933 | 0.689 | 0.475 | 1.894 |
| COPD | 0.627 | 0.393 | 2.344 | 0.634 | 0.403 | 1.822 | 0.653 | 0.426 | 1.810 |
| Arthritis | 0.818 | 0.668 | 2.737 | 0.814 | 0.662 | 2.754 | 0.866 | 0.750 | 2.442 |
| High Cholesterol | 0.706 | 0.500 | 2.163 | 0.695 | 0.483 | 2.298 | 0.763 | 0.582 | 2.739 |
| Kidney Disease | 0.074 | 0.006 | 0.967 | 0.272 | 0.074 | 0.955 | 0.429 | 0.184 | 0.764 |
| Heart Disease | 0.610 | 0.373 | 1.199 | 0.589 | 0.347 | 0.913 | 0.683 | 0.466 | 2.661 |
| Depression | 0.495 | 0.245 | 2.319 | 0.458 | 0.210 | 2.363 | 0.643 | 0.413 | 2.328 |
| Stroke | 0.541 | 0.293 | 1.117 | 0.489 | 0.239 | 0.948 | 0.535 | 0.286 | 0.881 |

*Table 10.  Percentage of county-level estimates from each SHAPE model and CDC PLACES that fall within the confidence intervals of the direct estimates across health outcomes. Values in parentheses indicate counts. \* Indicates that the estimates were generated without health risk behaviors (smoking and obesity) as predictor variables. \*\* Direct estimates dataset missing one value; totals based on 66 counties instead of 67.*

| SAE Outcome | Stratified Model | State-Specific Model | CDC PLACES |
|---|---|---|---|
| **New York (2021)** | | | |
| **Smoking*** | 79.03 (49) | 64.52 (40) | 66.13 (41) |
| **Obesity*** | 62.90 (39) | 62.90 (39) | 96.75 (60) |
| **Cancer** | 82.26 (51) | 91.94 (57) | 93.55 (58) |
| **Asthma** | 80.65 (50) | 75.81 (47) | 83.87 (52) |
| **High Blood Pressure** | 77.42 (48) | 85.48 (53) | 93.55 (58) |
| **Diabetes** | 90.32 (56) | 83.87 (52) | 93.55 (58) |
| **COPD** | 77.41 (48) | 64.52 (40) | 80.65 (50) |
| **Arthritis** | 74.19 (46) | 72.58 (45) | 98.39 (61) |
| **High Cholesterol** | 82.26 (51) | 80.65 (50) | 100 (62) |
| **Kidney Disease** | 24.19 (15) | 30.65 (19) | 27.42 (17) |
| **Heart Disease** | 79.03 (49) | 82.26 (51) | 0 (0) |
| **Depression** | 77.41 (48) | 70.97 (44) | 77.42 (48) |
| **Stroke** | 62.90 (39) | 79.03 (49) | 75.81 (47) |
| **Florida (2019)** | | | |
| **Smoking*** | 53.73 (36) | 61.19 (41) | 71.64 (48) |
| **Obesity*** | 55.22 (37) | 58.21 (39) | 67.16 (45) |
| **Cancer** | 61.19 (41) | 52.24 (35) | 80.60 (54) |
| **Asthma** | 76.12 (51) | 86.57 (58) | 85.07 (57) |
| **High Blood Pressure**\*\* | 66.67 (50) | 69.70 (46) | 74.24 (49) |
| **Diabetes** | 74.63 (50) | 77.61 (52) | 77.61 (52) |
| **COPD** | 61.19 (41) | 70.15 (47) | 67.16 (45) |
| **Arthritis** | 67.16 (45) | 70.15 (47) | 71.64 (48) |
| **High Cholesterol** | 82.09 (55) | 77.61 (52) | 77.61 (52) |
| **Kidney Disease** | 92.54 (62) | 80.60 (54) | 91.05 (61) |
| **Heart Disease** | 80.60 (54) | 82.09 (55) | 0 (0) |
| **Depression** | 77.61 (52) | 77.61 (52) | 77.61 (52) |
| **Stroke** | 76.12 (51) | 88.06 (59) | 89.55 (60) |

### G. Evaluating SHAPE Estimates Against CDC PLACES Estimates

*Table 11. Evaluation metrics comparing county-level SHAPE estimates against CDC PLACES estimates, by health outcome. * Indicates that the estimates were generated without health risk behaviors (smoking and obesity) as predictor variables.*

| | **Stratified Model** | | | **State-Specific Model** | | |
|---|---|---|---|---|---|---|
| **SAE Outcome** | r | $R^2$ | MAE | r | $R^2$ | MAE |
| **New York (2021)** | | | | | | |
| *Smoking** | 0.882 | 0.778 | 1.812 | 0.895 | 0.801 | 1.062 |
| *Obesity** | 0.707 | 0.500 | 4.259 | 0.733 | 0.537 | 3.786 |
| *Cancer* | 0.935 | 0.875 | 1.741 | 0.940 | 0.883 | 1.499 |
| *Asthma* | 0.381 | 0.145 | 1.175 | 0.700 | 0.490 | 0.617 |
| *High Blood Pressure* | 0.869 | 0.755 | 2.466 | 0.853 | 0.727 | 1.591 |
| *Diabetes* | 0.885 | 0.783 | 0.560 | 0.756 | 0.572 | 1.081 |
| *COPD* | 0.925 | 0.856 | 0.408 | 0.916 | 0.838 | 0.891 |
| *Arthritis* | 0.904 | 0.817 | 1.618 | 0.899 | 0.808 | 2.056 |
| *High Cholesterol* | 0.666 | 0.444 | 3.081 | 0.724 | 0.524 | 2.083 |
| *Kidney Disease* | 0.757 | 0.574 | 0.242 | 0.734 | 0.539 | 0.403 |
| *Heart Disease* | 0.923 | 0.852 | 2.056 | 0.917 | 0.842 | 2.301 |
| *Depression* | 0.862 | 0.744 | 1.108 | 0.837 | 0.701 | 1.596 |
| *Stroke* | 0.786 | 0.618 | 0.291 | 0.918 | 0.842 | 0.282 |
| **Florida (2019)** | | | | | | |
| *Smoking** | 0.924 | 0.854 | 2.997 | 0.930 | 0.865 | 1.908 |
| *Obesity** | 0.745 | 0.555 | 3.908 | 0.755 | 0.569 | 3.472 |
| *Cancer* | 0.928 | 0.862 | 1.840 | 0.926 | 0.857 | 2.205 |
| *Asthma* | -0.353 | 0.125 | 1.081 | 0.262 | 0.069 | 0.703 |
| *High Blood Pressure* | 0.931 | 0.867 | 1.434 | 0.933 | 0.871 | 1.412 |
| *Diabetes* | 0.886 | 0.786 | 1.143 | 0.847 | 0.717 | 0.866 |
| *COPD* | 0.893 | 0.797 | 2.080 | 0.916 | 0.839 | 0.718 |
| *Arthritis* | 0.946 | 0.895 | 2.184 | 0.929 | 0.863 | 1.567 |
| *High Cholesterol* | 0.851 | 0.724 | 2.354 | 0.936 | 0.875 | 2.565 |
| *Kidney Disease* | 0.528 | 0.279 | 0.552 | 0.814 | 0.662 | 0.340 |
| *Heart Disease* | 0.881 | 0.777 | 3.607 | 0.885 | 0.784 | 2.888 |
| *Depression* | 0.741 | 0.550 | 1.066 | 0.817 | 0.668 | 1.009 |
| *Stroke* | 0.939 | 0.883 | 0.804 | 0.944 | 0.891 | 0.209 |

Table 12. Evaluation metrics comparing census tract-level SHAPE estimates against CDC PLACES estimates, by health outcome. * Indicates that the estimates were generated without health risk behaviors (smoking and obesity) as predictor variables.

|  | Stratified Model | | | State-Specific Model | | |
|---|---|---|---|---|---|---|
| **SAE Outcome** | **r** | **R²** | **MAE** | **r** | **R²** | **MAE** |
| **New York (2021)** | | | | | | |
| *Smoking** | 0.768 | 0.590 | 3.122 | 0.672 | 0.452 | 3.084 |
| *Obesity** | 0.699 | 0.488 | 7.343 | 0.731 | 0.534 | 5.528 |
| *Cancer* | 0.792 | 0.627 | 0.991 | 0.824 | 0.678 | 0.932 |
| *Asthma* | 0.543 | 0.295 | 1.267 | 0.767 | 0.588 | 0.966 |
| *High Blood Pressure* | 0.778 | 0.605 | 5.500 | 0.760 | 0.577 | 4.524 |
| *Diabetes* | 0.771 | 0.595 | 2.037 | 0.766 | 0.586 | 2.330 |
| *COPD* | 0.624 | 0.389 | 0.564 | 0.692 | 0.478 | 1.378 |
| *Arthritis* | 0.732 | 0.536 | 1.136 | 0.779 | 0.607 | 4.551 |
| *High Cholesterol* | 0.792 | 0.627 | 4.952 | 0.734 | 0.539 | 2.084 |
| *Kidney Disease* | 0.745 | 0.555 | 2.189 | 0.741 | 0.549 | 0.474 |
| *Heart Disease* | 0.683 | 0.467 | 1.426 | 0.662 | 0.438 | 1.793 |
| *Depression* | 0.714 | 0.510 | 2.370 | 0.799 | 0.638 | 2.069 |
| *Stroke* | 0.771 | 0.594 | 0.868 | 0.604 | 0.365 | 0.627 |
| **Florida (2019)** | | | | | | |
| *Smoking** | 0.869 | 0.755 | 3.245 | 0.599 | 0.358 | 3.821 |
| *Obesity** | 0.733 | 0.537 | 5.036 | 0.653 | 0.427 | 4.009 |
| *Cancer* | 0.887 | 0.788 | 1.524 | 0.854 | 0.730 | 1.832 |
| *Asthma* | 0.418 | 0.174 | 1.405 | 0.330 | 0.109 | 1.064 |
| *High Blood Pressure* | 0.832 | 0.693 | 3.649 | 0.810 | 0.657 | 3.640 |
| *Diabetes* | 0.783 | 0.613 | 1.842 | 0.751 | 0.565 | 1.990 |
| *COPD* | 0.789 | 0.623 | 1.766 | 0.654 | 0.428 | 1.744 |
| *Arthritis* | 0.856 | 0.733 | 3.568 | 0.862 | 0.743 | 3.325 |
| *High Cholesterol* | 0.852 | 0.726 | 2.159 | 0.867 | 0.751 | 2.394 |
| *Kidney Disease* | 0.723 | 0.523 | 0.538 | 0.805 | 0.648 | 0.592 |
| *Heart Disease* | 0.786 | 0.619 | 2.680 | 0.790 | 0.624 | 2.114 |
| *Depression* | 0.614 | 0.377 | 1.675 | 0.645 | 0.417 | 1.989 |
| *Stroke* | 0.797 | 0.635 | 0.824 | 0.725 | 0.525 | 0.772 |

H. **Composite Scores and Model Ranking Results**

*Table 13. Calculated deviations for each evaluation metric used in the study's model ranking analysis * Indicates that the estimates were generated without health risk behaviors (smoking and obesity) as predictor variables.*

|  | Stratified Model | | | State-Specific Model | | | CDC PLACES | | |
|---|---|---|---|---|---|---|---|---|---|
| **SAE Outcome** | r Dev. | MAE Dev. | CI Dev. | r Dev. | MAE Dev. | CI Dev. | r Dev. | MAE Dev. | CI Dev. |
| **New York - 2021** | | | | | | | | | |
| *Smoking** | 0.143 | 0.143 | 0.060 | 0.149 | 0.146 | -0.085 | 0.235 | 0.203 | -0.069 |
| *Obesity** | -0.079 | -0.440 | -0.101 | -0.027 | -0.393 | -0.101 | 0.224 | 0.193 | 0.238 |
| *Cancer* | -0.258 | 0.328 | 0.093 | -0.246 | 0.361 | 0.189 | -0.236 | 0.486 | 0.206 |
| *Asthma* | -0.462 | 0.240 | 0.077 | -0.258 | 0.327 | 0.028 | 0.001 | 0.356 | 0.109 |
| *High Blood Pressure* | 0.066 | -0.005 | 0.044 | 0.047 | 0.127 | 0.125 | 0.292 | 0.350 | 0.206 |
| *Diabetes* | -0.012 | 0.383 | 0.173 | 0.017 | 0.350 | 0.109 | 0.115 | 0.436 | 0.206 |
| *COPD* | 0.089 | 0.406 | 0.044 | 0.035 | 0.351 | -0.085 | 0.110 | 0.418 | 0.077 |
| *Arthritis* | 0.171 | 0.125 | 0.012 | 0.169 | 0.095 | -0.004 | 0.312 | 0.338 | 0.254 |
| *High Cholesterol* | -0.259 | -0.007 | 0.093 | -0.225 | 0.104 | 0.077 | 0.086 | 0.334 | 0.270 |
| *Kidney Disease* | 0.033 | -0.085 | -0.488 | -0.043 | -0.057 | -0.424 | -0.274 | -0.109 | -0.456 |
| *Heart Disease* | -0.231 | 0.506 | 0.060 | -0.229 | 0.529 | 0.093 | -0.171 | 0.198 | -0.730 |
| *Depression* | 0.177 | 0.124 | 0.044 | 0.141 | 0.025 | -0.020 | 0.260 | 0.233 | 0.044 |
| *Stroke* | -0.407 | 0.504 | -0.101 | -0.252 | 0.544 | 0.060 | -0.255 | 0.537 | 0.028 |
| **Florida - 2019** | | | | | | | | | |
| *Smoking** | 0.101 | 0.059 | -0.193 | 0.154 | 0.170 | -0.118 | 0.237 | 0.245 | -0.014 |
| *Obesity** | 0.190 | -0.193 | -0.178 | 0.175 | -0.090 | -0.148 | 0.284 | 0.214 | -0.058 |
| *Cancer* | 0.119 | 0.296 | -0.118 | 0.108 | 0.234 | -0.208 | 0.172 | 0.490 | 0.076 |
| *Asthma* | -0.767 | 0.348 | 0.031 | -0.687 | 0.375 | 0.136 | -0.124 | 0.416 | 0.121 |
| *High Blood Pressure* | 0.187 | 0.193 | -0.063 | 0.188 | 0.191 | -0.033 | 0.267 | 0.258 | 0.012 |
| *Diabetes* | -0.018 | 0.286 | 0.016 | 0.055 | 0.334 | 0.046 | 0.112 | 0.342 | 0.046 |
| *COPD* | 0.050 | 0.255 | -0.118 | 0.057 | 0.356 | -0.029 | 0.076 | 0.358 | -0.058 |
| *Arthritis* | 0.241 | 0.179 | -0.058 | 0.237 | 0.176 | -0.029 | 0.289 | 0.236 | -0.014 |
| *High Cholesterol* | 0.129 | 0.290 | 0.091 | 0.118 | 0.264 | 0.046 | 0.186 | 0.179 | 0.046 |
| *Kidney Disease* | -0.503 | 0.521 | 0.195 | -0.305 | 0.523 | 0.076 | -0.148 | 0.560 | 0.181 |
| *Heart Disease* | 0.033 | 0.476 | 0.076 | 0.012 | 0.531 | 0.091 | 0.106 | 0.194 | -0.730 |
| *Depression* | -0.082 | 0.260 | 0.046 | -0.119 | 0.251 | 0.046 | 0.066 | 0.258 | 0.046 |
| *Stroke* | -0.036 | 0.492 | 0.031 | -0.088 | 0.524 | 0.151 | -0.042 | 0.537 | 0.166 |

***Average Values Used to Calculate Deviations***

*Average Pearson's r = 0.577*

*Average MAE = 0.439*

*Average CI Coverage = 0.730*

*Table 14. Calculated composite scores, ordered greatest to least, used for the models' ranking analysis * Indicates that the estimates were generated without health risk behaviors (smoking and obesity) as predictor variables.*

| SHAPE: Stratified Model | | SHAPE: State-Specific Model | | CDC PLACES | |
|---|---|---|---|---|---|
| **Health Measure** | **Score** | **Health Measure** | **Score** | **Health Measure** | **Score** |
| *Heart Disease* | 0.92 | *Heart Disease* | 1.03 | *Arthritis* | 1.42 |
| *Diabetes* | 0.83 | *Stroke* | 0.94 | *High Blood Pressure* | 1.39 |
| *COPD* | 0.73 | *Diabetes* | 0.91 | *Diabetes* | 1.26 |
| *Arthritis* | 0.67 | *COPD* | 0.69 | *Cancer* | 1.19 |
| *Depression* | 0.57 | *Arthritis* | 0.65 | *High Cholesterol* | 1.10 |
| *Stroke* | 0.48 | *High Blood Pressure* | 0.64 | *Obesity** | 1.09 |
| *Cancer* | 0.46 | *Cancer* | 0.44 | *COPD* | 0.98 |
| *High Blood Pressure* | 0.42 | *Smoking** | 0.42 | *Stroke* | 0.97 |
| *High Cholesterol* | 0.34 | *High Cholesterol* | 0.38 | *Depression* | 0.91 |
| *Smoking** | 0.31 | *Depression* | 0.32 | *Asthma* | 0.88 |
| *Kidney Disease* | -0.33 | *Asthma* | -0.08 | *Smoking** | 0.84 |
| *Asthma* | -0.53 | *Kidney Disease* | -0.23 | *Kidney Disease* | -0.25 |
| *Obesity** | -0.80 | *Obesity** | -0.58 | *Heart Disease* | -1.13 |

I. <u>**NHIS**</u> **Analysis: Evaluating SHAPE Estimates Against Direct Estimates*** *from BRFSS*

Table 15. Evaluation metrics comparing county-level SHAPE estimates **generated using NHIS as an input survey (instead of BRFSS)** against direct estimates from state-specific BRFSS surveys. * Indicates that the estimates were generated without health risk behaviors (smoking and obesity) as predictor variables. ** Indicates that SEER cancer registry data was used for evaluation instead of BRFSS direct estimates.

| | SHAPE: Stratified Model | | | CDC PLACES | | |
|---|---|---|---|---|---|---|
| SAE Outcome | r | $R^2$ | MAE | r | $R^2$ | MAE |
| New York (2021) | | | | | | |
| Smoking* | 0.691 | 0.478 | 2.842 | 0.812 | 0.659 | 2.615 |
| Obesity* | 0.467 | 0.218 | 4.876 | 0.801 | 0.642 | 2.665 |
| Cancer** | 0.777 | 0.604 | 2.655 | 0.783 | 0.613 | 1.416 |
| Asthma | 0.381 | 0.145 | 2.793 | 0.578 | 0.334 | 1.822 |
| High Blood Pressure | 0.581 | 0.337 | 3.826 | 0.869 | 0.755 | 1.853 |
| COPD | 0.581 | 0.338 | 1.864 | 0.687 | 0.472 | 1.501 |
| Diabetes | 0.441 | 0.351 | 1.834 | 0.692 | 0.479 | 1.408 |
| Florida (2019) | | | | | | |
| Smoking* | 0.745 | 0.555 | 2.911 | 0.814 | 0.663 | 2.393 |
| Obesity* | 0.783 | 0.613 | 4.682 | 0.861 | 0.741 | 2.555 |
| Cancer** | 0.762 | 0.581 | 1.128 | 0.749 | 0.561 | 1.125 |
| Asthma | 0.142 | 0.020 | 1.597 | 0.453 | 0.205 | 1.508 |
| High Blood Pressure | 0.732 | 0.536 | 2.964 | 0.844 | 0.713 | 2.327 |
| Diabetes | 0.629 | 0.396 | 2.489 | 0.689 | 0.475 | 1.894 |
| COPD | 0.609 | 0.371 | 3.649 | 0.653 | 0.426 | 1.810 |